\documentclass[journal]{new-aiaa} 
\usepackage[utf8]{inputenc}

\usepackage{graphicx}
\usepackage{amsmath}
\usepackage[version=4]{mhchem}
\usepackage{siunitx}
\usepackage{longtable,tabularx}
\usepackage{subcaption}

\newcommand{\bettershortstack}[4][c]{%
  \renewcommand{\arraystretch}{#2}
  \begin{tabular}[b]{@{}#1@{}}
  #4
  \end{tabular}%
  \renewcommand{\arraystretch}{#3}
}

\title{Three-Dimensional Kinetic Simulation of an Ion Thruster Plume with Carbon Backsputtering in a Vacuum Chamber}

\author{Keita Nishii\footnote{Postdoctoral Researcher, Department of Aerospace Engineering, University of Illinois, Urbana-Champaign, Member AIAA.} and Deborah A. Levin\footnote{Professor, Department of Aerospace Engineering, University of Illinois, Urbana-Champaign, Fellow AIAA.}}

\begin{document}

\maketitle



\section*{Abstract}


\begin{abstract}
Gridded ion thrusters are tested in ground vacuum chambers to verify their performance when deployed in space. However, the presence of high background pressure and conductive walls in the chamber leads to facility effects that increase uncertainty in the performance of the thruster in space. To address this issue, this study utilizes a fully kinetic simulation to investigate the facility effects on the thruster plume. The in-chamber condition shows a downstream neutral particle density 100 times larger than the in-space case due to ion neutralization at the wall and limited vacuum pump capability, resulting in a significant difference in the density and distribution of charge-exchange ions. The flux, energy, and angle of charge-exchange ions incident on the chamber wall are found to be altered by the electron sheath, which can only be simulated by the fully kinetic approach, as opposed to the conventionally used quasi-neutral Boltzmann approach. We also examine the effect of backsputtering, another important facility effect, and find that it does not necessarily require a fully kinetic simulation as the incident flux and energy of the sampled charge-exchange ion are negligibly small. Finally, we demonstrate that the carbon deposition rate on the thruster is significantly influenced by the angular dependence of the sputtered carbon, with a nearly 50\% effect.
\end{abstract}

\section*{Nomenclature}


{\renewcommand\arraystretch{1.0}
\noindent\begin{longtable*}{@{}l @{\quad=\quad} l@{}}
$E$                         & electric field, \si{V \cdot m^{-1}} \\
$E_\mathrm{B}$              & binding energy of carbon atoms, \si{eV} \\
$E_\mathrm{b}$              & kinetic energy of a beam ion, \si{eV} \\
$E_\mathrm{c}$              & kinetic energy of a sputtered carbon particle, \si{eV} \\
$E_\mathrm{i}$              & kinetic energy of an incident ion, \si{eV} \\
$e$                         & elementary charge, \si{C} \\
$F_\mathrm{col}$            & DSMC module interval \\
$F_\mathrm{num}$            & superparticle factor \\
$f$                         & probability function \\
$g$                         & relative speed of colliding particles, \si{m \cdot s^{-1}} \\
$j_\mathrm{i}$              & incident ion particle flux on wall, \si{m^{-2} \cdot s^{-1}}\\
$j_\mathrm{n,ex}$           & thruster exit particle fluxes of Xe and Xe$^+$, \si{m^{-2} \cdot s^{-1}}\\
$j_\mathrm{total}$          & total fluxes of heavy particles, \si{m^{-2} \cdot s^{-1}} \\
$k_\mathrm{B}$              & Boltzmann constant, \si{m^{2} \cdot kg \cdot s^{-2}} \\
$L/D$                       & aspect ratio of the vacuum chamber \\
$m$                         & interatomic potential exponent parameter \\
$m_\mathrm{e}$              & electron mass, \si{kg} \\
$n_\mathrm{ex}$             & number density at the thruster exit, \si{m^{-3}} \\
$n_\mathrm{n,i,e,CEX,c}$    & number densities of Xe, Xe$^+$, e$^-$, Xe$^+_\mathrm{CEX}$, and C, \si{m^{-3}} \\
$n_\mathrm{0}$              & reference number density, \si{m^{-3}} \\
$R$                         & thruster exit radius, \si{m} \\
$r$                         & radial coordinate, \si{m} \\
$T_\mathrm{e0}$             & reference electron temperature, \si{K} \\
$W_\mathrm{n,i}$            & species weighting factors of Xe and Xe$^+$ \\
$x$, $y$, $z$               & Cartesian coordinates, \si{m} \\
$x_0$, $y_0$, $z_0$         & center positions of the thruster exit in ($x$, $y$, $z$), \si{m} \\
$Y$                         & total sputter yield \\
$Y_\mathrm{D}$              & differential sputter yield \\
$Y_0$                       & normal incident sputter yield \\
$Y'$                        & angular dependence of sputter yield \\
$\alpha$                    & polar emitted angle of a sputtered carbon particle, \si{deg} \\
$\beta$                     & azimuthal emitted angle of a sputtered carbon particle, \si{deg} \\
$\Delta t_\mathrm{n,i,e}$   & timesteps of Xe, Xe$^+$ and e$^-$, \si{s} \\
$\Delta x$                  & cell size, \si{m} \\
$\varepsilon_0$             & permittivity of free space, \si{F \cdot m^{-1}} \\
$\eta_\mathrm{u}$           & propellant utilization efficiency \\
$\theta$                    & polar angle of an incident ion, \si{deg} \\
$\lambda$                   & mean free path, \si{m} \\
$\lambda_\mathrm{D}$        & Debye length, \si{m} \\
$\rho$                      & volume charge density, \si{C \cdot m^{-3}} \\ 
$\sigma_\mathrm{CEX}$       & charge exchange collision cross-section, \si{m^{2}} \\
$\phi$                      & electric potential, \si{V} \\
$\phi_0$                    & reference electric potential, \si{V} \\
$\Omega$                    & solid angle, \si{sr} \\
$\omega_\mathrm{pe}$        & plasma frequency, \si{s^{-1}} \\

\end{longtable*}}


\section{Introduction}

The use of electric propulsion (EP) on satellites has been increasing in recent decades~\cite{Lev2019-jv}. In particular, gridded ion thrusters (GIT) have been widely developed and used for their high specific impulse and high efficiency~\cite{Holste2020-gr}. Ground facility testing is essential to estimate their performance and lifetime for actual missions. However, there are several limitations in ground testing~\cite{Foster2022-ck}. Typical operating chamber pressures are approximately three orders of magnitude higher than the ambient gas pressures of 10$^{-9}$ Torr on orbit (at 1000 km). At such high background pressure, the lifetime of ion grid optics is impacted by erosion driven by charge exchange (CEX) ions. Another limitation is the surrounding wall of the vacuum chamber. The high energy ion bombardment on the wall sputters wall material, and its deposition masks erosion caused by CEX backflow ions. In addition, ion beam neutralization depends on the neutralizer to ion beam plasma bridge, which can be affected by nearby wall surfaces and CEX collisions. For reasons such as these, it is important to understand the interaction between the thruster plasma plume and the ground facility.

Numerical simulation, such as Particle-in-Cell (PIC)~\cite{Birdsall1991-lh} and Direct Simulation Monte Carlo (DSMC)~\cite{Bird1987-ba}, is a strong tool to isolate facility effects from on-orbit performance. Several computational studies investigate the EP thruster plume in space and how the plume interacts with the spacecraft via CEX ions~\cite{Jambunathan2020-io, Nuwal2020-pa, Tajmar2001-qe, Wang2001-sc, Korkut2017-wg, Korkut2017-zq, Cai2018-dj, Araki2019-gt, Kang2023-ti, Roy1996-ub, Choi2008-ca, Cai2015-fq, Wang2019-gy}. Most previous studies of the EP thruster plumes do not model electrons as particles and assumed a Boltzmann or polytropic relationship for electron temperature. Among them, some studies assumed quasi-neutrality~\cite{Tajmar2001-qe, Wang2001-sc, Korkut2017-wg, Korkut2017-zq, Cai2018-dj, Araki2019-gt, Kang2023-ti}, and other studies use a fluid approach for electrons (hybrid-PIC)~\cite{Roy1996-ub, Choi2008-ca, Cai2015-fq, Wang2019-gy}. However, the hybrid-PIC approach cannot even resolve ion beam neutralization close to the thruster and electron temperature in the transverse direction since the electrons are nonequilibrium and anisotropic in nature~\cite{Wang2019-gy}. Our previous studies showed that the energy of CEX ions incident on the solar array panels differs by one order of magnitude between the Boltzmann and fully kinetic results~\cite{Jambunathan2020-io, Nuwal2020-pa}, which also indicates the importance of the fully kinetic approach for vacuum chamber modeling.

A few studies have investigated the CEX ion population in a vacuum chamber, which has relatively high background pressure. Korkut et al.~\cite{Korkut2017-wg} investigated the difference in the CEX ion distribution between the in-space and in-chamber conditions for several GIT configurations using a quasi-neutral Boltzmann approach. The study showed that the CEX ion number density increased as the background neutral number density increased by changing the vacuum pump size. Cai~\cite{Cai2015-fq} modeled the ground operation of a cluster of BHT-200 Hall thrusters using a hybrid-PIC approach. The study captured the cluster effect, including CEX ions created by finite background pressure. Unlike the above studies, Wang et al.~\cite{Wang2015-pm} simulated beam ion plumes inside a vacuum chamber using a fully kinetic approach, which directly models an electron as a particle. The study showed the difference in the electric potential in the vacuum chamber from in space and the CEX ion distribution in the vacuum chamber using the overlay approach. However, their ion source size, gas species, and propellant utilization efficiency were different from typical conditions of GITs that have actually been operated in space. Therefore, further efforts are required to investigate the EP thruster plume interaction with the vacuum chamber.

As for the facility effect, not only the background pressure effect but also the backsputtering is another main focus of this study. Backsputtering is one of the largest concerns for higher-power GIT testing~\cite{Polk2000-dk, Hickman2005-ha, Williams2005-yn}. Although EP testing facilities are lined with graphite to reduce the backsputtering caused by the high-energy ion beam impingement, investigations demonstrate that deposition on the thruster still occurs, resulting in high uncertainties in the measurement of both thruster component erosion rates and the proportion of deposited carbon films associated with thruster erosion~\cite{Jovel2022-tu}. Some numerical studies have modeled backsputtered carbon to identify how much backsputtered carbon deposits on the thruster~\cite{Gilland2016-ed, Zheng2017-uz, Choi2017-xe}. Gilland et al.\cite{Gilland2016-ed} calculated the sputtered carbon using the DSMC method using a Hall thruster plume model numerically obtained in a different study~\cite{Lopez_Ortega2015-fw}. However, the simulated deposition rate was five times larger than that experimentally observed. Zheng et al.~\cite{Zheng2017-uz} also calculated sputtered carbon using an analytical GIT thruster plume and found the effective beam target geometry required to reduce the backsputtering. Choi et al.~\cite{Choi2017-xe} numerically simulated the near-thruster region to understand the origins of carbon deposited on the inner wall of the discharge channel, e.g., carbon sputtering of the ground facility, the inner pole cover, or the outer pole cover.

Although a number of efforts have been made to investigate the facility effect, there is no simulation for an entire vacuum chamber with the fully kinetic PIC approach. Furthermore, no study has simulated backsputtering and the thruster plume simultaneously. The former simulation is important for clarifying how the exact electric field solved by the fully kinetic approach differs from the commonly used Boltzmann approach and, ultimately, to know how it affects the facility plasma. The latter is also important to get a more explicit understanding of how the thruster plume in the vacuum chamber causes sputtering. Therefore, through three-dimensional fully kinetic electrostatic simulations, this study seeks to understand 1) the difference between the GIT plasma plume in space and in the vacuum chamber, 2) the impact of the Boltzmann assumption on the facility plasma, 3) the effect of the thruster plume density on the facility plasma, and 4) the influence of the ion-surface interaction model on carbon backsputtering.

Solving a fully kinetic simulation of an actual EP plume, which has a number density of $\sim 10^{15}$ \si{ m^{-3}} in a 1-meter scale computational geometry, is presently not possible with current computer resources. Therefore, many previous fully kinetic EP plume simulations have either used extremely small thrusters (about 10 times the Debye length) or scaling techniques~\cite{Wang2015-pm, Wang2019-gy}. However, dimensional scaling does not scale with the Debye length, and reducing ion-to-electron mass ratio greatly influences plasma properties~\cite{Yuan2020-ma}. We calculate the actual GIT operation geometry using the real ion-to-electron (Xe$^+$/e$^-$) mass ratio without dimensional scaling using our in-house 3-D PIC-DSMC solver, Cuda-based Hybrid Approach for Octree Simulations (CHAOS)~\cite{Jambunathan2018-xf}, which has MPI-CUDA parallelization strategies. This study selects maximum thruster density as $4\times 10^{14}$ \si{ m^{-3}}, which is one order of magnitude lower than that of GITs commonly used in space due to computational limitations. However, the investigation of the density effect, the third objective of this study, allows us to predict what would happen at higher densities. In addition, since this density is only a few times smaller than the actual thruster density at the minimum throttle level~\cite{Roy1996-lu, Takao2006-qp}, the result is comparable to an actual ground firing experiment.

The outline of this article is as follows. Section~\ref{sec:numericalapproach} reviews our plasma modeling approach and describes the ion-surface interaction model newly implemented in CHAOS. Section~\ref{sec:conditions} explains the different conditions, such as geometry, electric field model, and plasma densities considered. Finally, in Section~\ref{sec:result}, the comparisons between the simulated cases are presented and analyzed based on the objectives of this study.


\section{Numerical Approach}
\label{sec:numericalapproach}

This section discusses the computational framework implemented in CHAOS to couple the PIC and DSMC approaches to calculate the self-consistent electric field. In the ion thruster plasma plume, the time and length scales that regulate the two primary physical processes, namely collisions and the electric field, differ by at least two orders of magnitude as a result of the variations in the particle number densities, velocities, and mass of the plume species. CHAOS has been extensively used for ion thruster plume simulations and has several computational techniques, which are described in our previous papers~\cite{Jambunathan2018-xf, Jambunathan2020-io, Jambunathan2020-yx, Nuwal2020-pa}. Here we briefly present an overview of the DSMC and PIC modules in Section \ref{subsec:dsmc-pic}. Section~\ref{subsec:pic-dsmccounple} discusses the numerical methodologies used to couple the DSMC and PIC approaches that model these physical processes with disparate time and length scales.

In addition, we have extended CHAOS to model ground facility environments in this work. There are two important plasma-surface interactions; charge absorption and sputtering. Most ions lose their positive charge and are neutralized when they hit the wall. Since an important objective of this study is to model sputtered particles from the vacuum chamber walls, Section~\ref{subsec:surface} discusses the surface interaction models used in this study.

\subsection{DSMC and PIC Modules}
\label{subsec:dsmc-pic}

The DSMC module models three types of collisions: momentum exchange (MEX) collisions between Xe-Xe and Xe-Xe$^+$, and CEX collisions between Xe-Xe$^+$. The collision cross sections for MEX between neutral particles and the MEX and CEX collisions between neutral particles and ions are obtained from Refs.~\cite{Araki2013-em}, and \cite{Miller2002-ne}, where the details of the collision scheme are described by Korkut et al.~\cite{Korkut2015-fo} and Serikov et al.~\cite{Serikov1999-jq}. The no-time-counter (NTC) collision scheme proposed by Serikov et al.~\cite{Serikov1999-jq} is used in this study since it accounts for the disparate timesteps and weighting factors of ions and neutral particles. The neutral particles move only when the DSMC module is executed since the velocity of neutral particles is updated only by the DSMC collisions and not by the electric field computed in the PIC module. In contrast, the ions and electrons move every iteration.

The main objective of the PIC module is to compute the self-consistent electric field based on the spatial distribution of the charged particles. We calculate electric potential by a fully kinetic approach as an explicit PIC technique~\cite{Jambunathan2018-xf} and a quasi-neutral approach~\cite{Jambunathan2020-io, Nuwal2020-pa} as well. In the fully kinetic approach, the electric field, $E$, is solved based on Poisson’s equation as follows:
\begin{equation}
    \label{eq:poisson}
    \nabla^2\phi = - \frac{\rho}{\varepsilon_0},
\end{equation}
\begin{equation}
    \label{eq:efield}
    E = - \nabla\phi,
\end{equation}
where $\phi$ is the electric potential, $\rho$ is the volume charge density and $\epsilon_0$ is the permittivity of free space in vacuum. As in our previous works~\cite{Jambunathan2018-xf, Jambunathan2020-io, Jambunathan2020-yx, Nuwal2020-pa}, a finite volume approach based on an unstructured octree grid is used to solve Eq.~\eqref{eq:poisson}. The electric grid satisfies the grid refinement criteria that the octree cell size must be smaller than the local Debye length.

Quasi-neutral assumptions, such as a Boltzmann model, are often used and be adequate in modeling the core region of the ion thruster plume~\cite{Korkut2015-fo, Korkut2017-zq}. In contrast to the fully kinetic approach, the quasi-neutral approach does not have cell size criteria based on the local Debye length, significantly reducing the computational cost. We also use the Boltzmann model to compare those obtained from the fully kinetic PIC-DSMC simulations. The electric potential for the Boltzmann case is computed as follows:
\begin{equation}
    \label{eq:boltz}
    \phi = \phi_0 + \frac{k_B T_\mathrm{e0}}{e}\ln{\left(\frac{n_\mathrm{e}}{n_{0}}\right)},
\end{equation}
where $\phi_0$ is the reference potential, $n_\mathrm{0}$ is the reference number density, $k_\mathrm{B}$ is the Boltzmann constant, $T_{e0}$ is the reference electron temperature, and $e$ is the elementary charge. The electron number density, $n_\mathrm{e}$, is obtained by invoking the quasi-neutrality assumption with $n_\mathrm{e} = n_\mathrm{i}$. In the Boltzmann cases, the electric field is calculated by Eq.~\eqref{eq:efield} based on the potential obtained by Eq.~\eqref{eq:boltz}.

\subsection{PIC-DSMC Coupling}
\label{subsec:pic-dsmccounple}

In this section, we briefly explain the PIC and DSMC coupling method used in CHAOS (see Ref.~\cite{Jambunathan2020-io} for more details). First, CHAOS uses a separate linearized forest of octrees and domain decomposition for the PIC and DSMC simulations. A forest of octrees (FOT) is constructed and linearized using a Morton Z-curve to store only the final leaf nodes obtained during the octree adaptive mesh refinement (AMR) process. We apply the adaptive mesh refinement method to the octree leaf nodes since the number density can vary widely in the computational domain for expanding jets and beams. In addition to the variation of the local length scales, the mean free path and Debye length differ by at least three orders of magnitude for ion thruster plumes. Typically, the Debye length is on the order of 0.1-1 mm, while the mean free path is larger than 1 m. Therefore, to resolve these disparate length scales without compromising computational efficiency, we use two linear FOTs. The FOT constructed to resolve the local mean free path criterion, $\Delta x<\lambda$, in the DSMC module is called the C-FOT, and the FOT constructed to resolve the local Debye length criterion, $\Delta x <\lambda_\mathrm{D}$ in the PIC module is called the E-FOT. This strategy of using two separate grids was first proposed by Serikov et al.~\cite{Serikov1999-jq}.

Second, weighting factors, $W$, are used to increase the number of charged computational particles compared with the neutral particles due to the disparate length scales of the C- and E-FOTs, and disparate number densities of the neutral particles and CEX ions. Third, the time-slicing of the DSMC and PIC modules and species-dependent timesteps is implemented due to the different timescales for collision and plasma frequencies. The positions of the neutral particles, ions, and electrons are updated with timesteps of $\Delta t_\mathrm{n} \gg \Delta t_\mathrm{i} = \Delta t_\mathrm{e}$ to reconcile these disparate timescales. In addition, the DSMC module is called every $F_\mathrm{col}$ PIC iteration to reduce the calculation cost when the timestep of ions and electrons is very small. This study uses $F_\mathrm{col}=100$ for fully kinetic simulations. 
Also, for modeling ion-neutral collisions, we use an additional relationship between timesteps, $F_\mathrm{col}$, and weighting factors to account for the large timesteps species connections~\cite{Korkut2015-fo}:
\begin{equation}
    \label{eq:neutionweight}
    \frac{\Delta t_\mathrm{i}\cdot F_\mathrm{col}}{\Delta t_\mathrm{n}}=\frac{W_\mathrm{i}}{W_\mathrm{n}}.
\end{equation}

Fourth, after the spatial variation of the electric field in the plume simulations reaches a steady state, the electric-field macroparameters are sampled. However, since the low-velocity CEX ions attracted toward the backflow region move slowly, they require many more timesteps and, thus, more number iterations to reach a steady state. For the thruster plume simulations performed in this article, the electric field reaches a steady state at nearly 20 \si{\micro s}, while the neutral and ion flux reaches a steady state at nearly 400 \si{\micro s}. The Poisson solver is not used after the electric field reaches a steady state to save computational time without compromising accuracy. Instead, the sampled steady-state electric field continues to accelerate the charged particles. This ensures that more DSMC collisions are sampled to obtain good statistics for MEX and CEX ions while also saving computational time that would have been used to sample the steady-state electric field. Such an approximation is acceptable because the number density of CEX ions is typically less than 1\% of the beam ion density and therefore does not significantly affect the electric field. Figure~\ref{fig:timehist} and Table~\ref{tab:samplingsteps} in the appendix section~\ref{subsec:convergence} explain this computational strategy in detail and show how the number of computed particles evolves.

\subsection{Ion-Surface Interaction Models}
\label{subsec:surface}

Neutralized ions affect the background pressure distribution because the conductive wall surface removes the charge from ions and neutralizes them. This modeling is especially important when the propellant utilization efficiency is greater than 0.5 since the flux of heavy particles as ions become dominant, where the propellant utilization efficiency, $\eta_\mathrm{u}$, is calculated as follows:
\begin{equation}
    \eta_\mathrm{u} = \frac{j_\mathrm{ex}}{j_\mathrm{total}},
    \label{eq:propeff}
\end{equation}
where $j_\mathrm{ex}$ and $j_\mathrm{total}$ are the particle fluxes of accelerated ions and total heavy particles exhausted from the thruster. We assume all incident ions are neutralized at the wall. When hitting the wall, a computational ion particle changes its type to a neutral particle and leaves the surface with speed accommodated to the wall temperature. When we change the type of species, the differences in timestep and species weight between neutral particles and ions must be considered. However, each incident ion corresponds to one neutral particle since Eq.~\eqref{eq:neutionweight} is satisfied. In the fully kinetic simulations, we assume all electrons are absorbed into the wall, which means that the computational electron particles are deleted when they reach the computational domain boundary.

Regarding sputtered carbon, we assume it is mono-atomic for simplicity, although the experimental result shows some carbon clusters (e.g., C$_2$ and C$_3$)~\cite{Oyarzabal2008-ud}. We also assume that the sticking coefficient of a carbon particle is unity, which means that the computational carbon particles are deleted from the calculation when they hit any walls. Carbon flow is assumed to be collisionless, i.e., carbon particles do not collide with Xe, Xe$^+$, or each other.

The total sputter yield is calculated according to the work of Yim~\cite{Yim2017-ps}, who surveyed low energy (< 1000 eV) xenon ion impact sputter yields to develop a coherent baseline set of sputter yield data fits for modeling electric propulsion systems. The total sputter yield, $Y(E_\mathrm{i},\theta)$, is calculated as a multiplicative factor of the energy-dependent normal sputter yield, $Y_0(E_\mathrm{i})$, and the angular dependence, $Y^\prime(\theta)$, as follows:
\begin{equation}
    \label{eq:totalyield}
    Y(E_\mathrm{i},\theta)= Y_0(E_\mathrm{i}) \cdot Y^\prime(\theta),
\end{equation}
where $E_\mathrm{i}$ is the total kinetic energy of the incident ion, and $\theta$ is the polar angle of the incident ion. Detailed expressions of $Y(E_\mathrm{i},\theta)$ and $Y_0(E_\mathrm{i})$ are given in Ref.~\cite{Yim2017-ps}.

\begin{figure}[hbt!]
\centering
\includegraphics[width=0.5\textwidth]{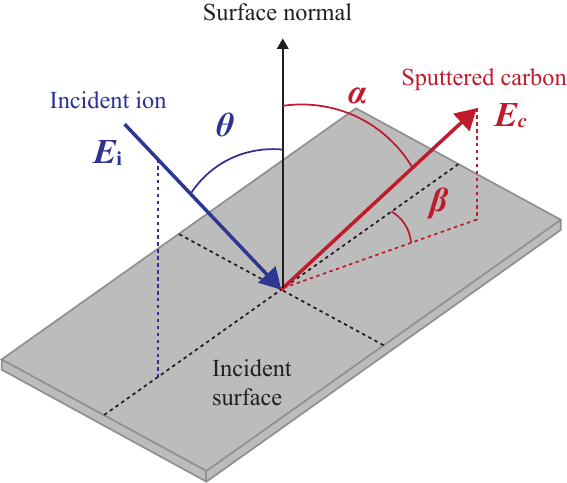}
\caption{Coordinate system defined for a sputtering model in this study.}
\label{fig:sputterschem}
\end{figure}

\begin{figure}[hbt!]
\centering
\includegraphics[width=0.5\textwidth]{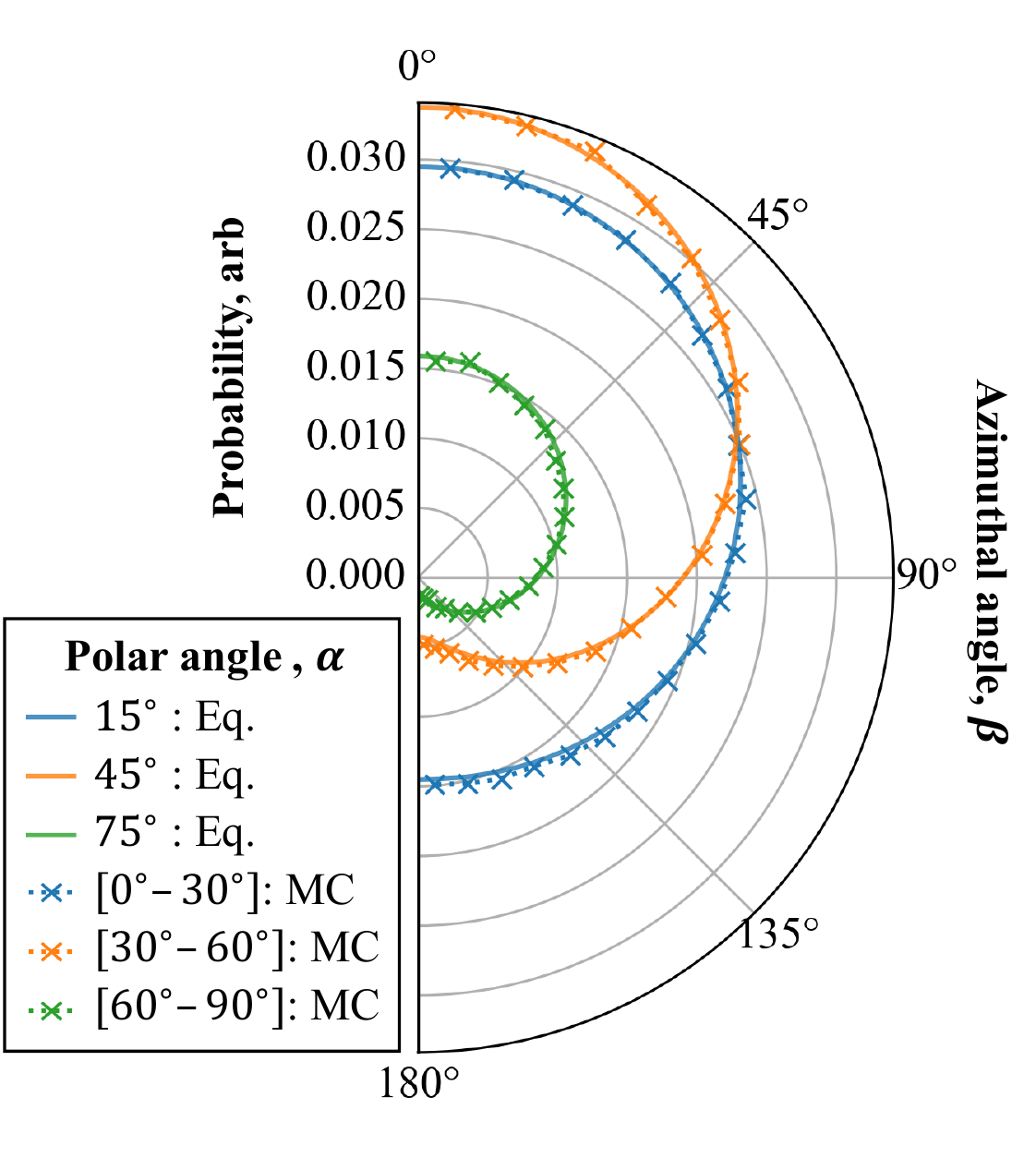}
\caption{Yim's~\cite{Yim2022-dj} differential sputtering yield of carbon when $E_\mathrm{i}=500$ eV and $\theta=30^\circ$ as an example. The solid lines (label: Eq.) show the original probability function, $f\left(E_\mathrm{i},\theta,\alpha,\beta\right)=Y_\mathrm{D}\left(E_\mathrm{i},\theta,\alpha,\beta\right)\sin\alpha$. The cross plots (label: MC) show the result of Monte Carlo sampling implemented in this study using the acceptance-rejection method.}
\label{fig:acceptreject}
\end{figure}

We compare two types of angular distribution: cosine distribution and a semi-empirical expression based on experimental measurements~\cite{Yim2022-dj}; the latter is denoted as Yim's model. Cosine distributions are commonly assumed in most sputtering models of electric propulsion devices~\cite{Van_Noord2005-kc, Gilland2016-ed, Choi2017-xe}. However, Yim~\cite{Yim2022-dj} presented the angular distribution of the differential yield, $Y_\mathrm{D}$, as a function of incident ion energy, $E_\mathrm{i}$, polar angle of an incident ion, $\theta$, polar angle of a sputtered carbon particle, $\alpha$, and azimuthal angle of sputtered carbon particle, $\beta$, as shown in the Fig~\ref{fig:sputterschem}. The detailed explanation of $Y_\mathrm{D}$ is given in Ref.~\cite{Yim2022-dj}. The total sputter yield, $Y(E_\mathrm{i},\theta)$, is also obtained from the solid angle integral of the differential yield, $Y_\mathrm{D}\left(E_\mathrm{i},\theta,\alpha,\beta\right)$, i.e.:
\begin{equation}
    Y(E_\mathrm{i},\theta) =  \int Y_\mathrm{D}\left(E_\mathrm{i},\theta,\alpha,\beta\right) \,d\Omega = \iint Y_\mathrm{D}\left(E_\mathrm{i},\theta,\alpha,\beta\right)\sin\alpha \,d\alpha\,d\beta.
\end{equation}
where $\Omega$ is a solid angle ($\Omega = \sin\alpha \cdot \alpha \cdot \beta$). In this work, we recover this differential sputter yield using the acceptance-rejection method~\cite{Robert2005-jx} to determine the sputtered angles of the surface carbon ($\alpha,\beta$) for the given incident Xe$^+$ energy, $E_\mathrm{i}$ and angle $\theta$. In this model, the probability function of the emission angle of a sputtered carbon atom, $f\left(E_\mathrm{i},\theta,\alpha,\beta\right)$, is $Y_\mathrm{D} \left(E_\mathrm{i},\theta,\alpha,\beta\right) \sin\alpha$.  Figure~\ref{fig:acceptreject} shows the differential sputtering yield of carbon when $_\mathrm{i} = 500$ eV and $\theta = 30^\circ$ as an example. The original probability function (label: Eq.) and the Monte Carlo sampling result in this study (label: MC) are in good agreement, which shows that the differential sputtering yield calculation has been successfully implemented.

The energy of the sputtered carbon from the surface, $E_\mathrm{c}$, is characterized using the Sigmund-Thompson energy distribution~\cite{Sigmund1981-se, Gnaser2007-bs}. The probability function, $f(E_\mathrm{c})$, is given as follows:
\begin{equation}
 f(E_\mathrm{c}) \propto \frac{E_\mathrm{c}}{\left(E_\mathrm{c}+E_\mathrm{B}\right)^{3-2m}},
\end{equation}
where $E_\mathrm{B}=7.4$ eV is the binding energy, and $m = \frac{1}{3}$ is the interatomic potential exponent parameter~\cite{Choi2017-xe}. This study assumed that the sputtered carbon energy follows this energy distribution function, with the maximum value being the energy of the colliding ions. This function reaches a maximum value when $E_\mathrm{c}=\frac{E_\mathrm{B}}{2(1-m)}$. Similar to the differential sputtering yield, we reproduce this probability function in our DSMC code using the acceptance-rejection method~\cite{Robert2005-jx}.

\section{Simulation Conditions}
\label{sec:conditions}

\subsection{Calculation Geometry}

\begin{figure}[hbt!]
\centering
\includegraphics[width=0.45\textwidth]{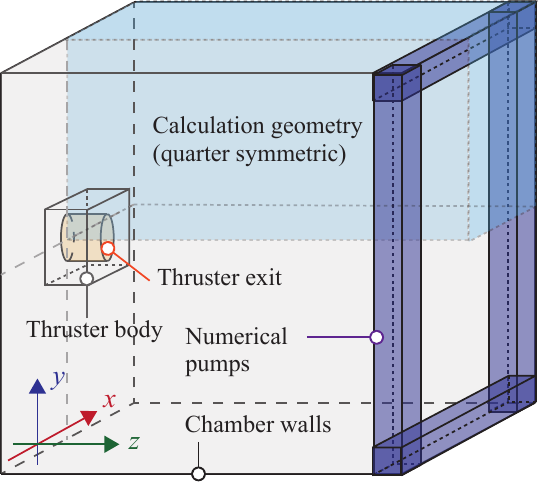}
\caption{Three-dimensional schematic of the target vacuum chamber geometry. This study only simulates a quarter domain (blue volume) due to symmetry.}
\label{fig:3Dschem}
\end{figure}

\begin{figure}[hbt!]
    \begin{subfigure}{0.39\textwidth}
        \centering
        \includegraphics[width=\linewidth]{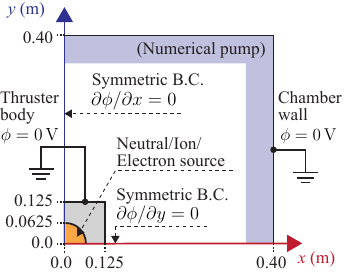}
        \caption{In $x$-$y$ plane} 
        \label{fig:2Dschem_xy}
    \end{subfigure}%
    \begin{subfigure}{0.61\textwidth}
        \centering
        \includegraphics[width=\linewidth]{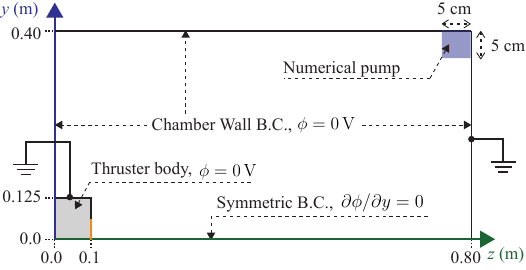}
        \caption{In $z$-$y$ plane} 
        \label{fig:2Dschem_zy}
    \end{subfigure}%
    \caption{Computational domain setups in two-dimensional planes. Symmetric BCs are applied on the $x=0$ m and $y=0$ m planes. Dirichlet and Neumann BCs are imposed on the outer boundary for the in-chamber and in-space cases, respectively.} 
    \label{fig:2Dschem}
\end{figure}

The target vacuum chamber geometry is shown in Fig.~\ref{fig:3Dschem} for an ion thruster installed in a cubic vacuum chamber. Unlike in our previous works~\cite{Jambunathan2020-io, Nuwal2020-pa}, this study only simulates a quarter domain due to symmetry to further save computational effort. Figure~\ref{fig:2Dschem} shows the simulation setup seen from the $x$-$y$ plane (a) and the $z$-$y$ plane (b). On the symmetric boundaries ($x=0$ m and $y=0$ m), the specular reflection boundary condition (BC) is implemented in the DSMC module, and the Neumann BC is implemented in the PIC module, i.e., $\partial\phi/\partial x=0$ or $\partial\phi/\partial y=0$. For the in-chamber conditions, a fully diffuse reflection condition is implemented on the wall BCs in the DSMC module, i.e., heavy particles are accommodated with the wall temperature of 300 K. As mentioned in Section~\ref{subsec:surface}, ions are neutralized when they reach the wall, and electrons are absorbed by the wall. The walls are grounded in the PIC module, meaning a Dirichlet BC of $\phi = 0$ V is implemented on the wall domain boundaries ($x=0.4$ m, $y=0.4$ m, $z=0$ m, and $z=0.8$ m planes). 

Numerical pumps, shown as the deep blue volume in Figs.~\ref{fig:3Dschem} and \ref{fig:2Dschem}, are introduced to discharge heavy particles from the vacuum chamber. Any computational particles entering this volume are deleted from the calculation, which is the same method used in our previous study~\cite{Korkut2017-zq}. The pump location is at the corner of the downstream wall so as not to disturb the plume expansion. The cross-section area of the numerical pump in the $z$-$y$ plane is $5 \times 5$ cm$^2$ to represent a typical vacuum chamber background pressure. The pumping speed of these numerical pumps is about 20 kL/s, which is the typical order of the pumping speed of a single cryopump~\cite{Spektor2021-jv}.

The center position of the cylindrical thruster exit with a radius of 0.0625 m is ($x_0,y_0,z_0$) = (0.0 m, 0.0 m, 0.1 m). The domain includes the thruster body with the thruster exit offset from the inlet plane at $z = 0.1$ m. The dimension of the thruster is the same as used in our previous calculation of an ion thruster system in space~\cite{Jambunathan2020-io, Nuwal2020-pa}. The boundary condition of the thruster body, including the thruster exit, is the same as the domain wall BC; diffuse reflection and 0 V Dirichlet electric potential BCs.

\subsection{Case Description}

We test five conditions covering different simulation geometries, electric potential model, and ion density at the thruster exit, as shown in Table~\ref{tab:caseid}. The in-space geometry simulation, designated as "0," is calculated to understand the difference between the space and chamber environments. Unlike the chamber geometry conditions, outer boundaries are modeled as a vacuum using charge-conserving energy-based (CCE) open BCs developed by Jambunathan and Levin~\cite{Jambunathan2020-yx}. The fully kinetic simulations are designated as "A," and the simulations using a quasi-neutral Boltzmann approach are designated as "B." The in-chamber simulations selected to study the effect of number density on the plume and facility plasma characteristics with thruster exit ion number densities of $4.0\times 10^{13}$ \si{m^{-3}} and $4.0\times 10^{14}$ \si{m^{-3}} are designated as "1" and "2," respectively.

\begin{table}[hbt!]
\caption{Test conditions of each case ID.}
\centering
\begin{tabular}{c|cccc}
\hline \hline
Case ID & Simulation geometry & E-field solver & Xe$^+$ number density, $n_0$    & Plasma frequency, $\omega_\mathrm{pe}$ \\  \hline
0A      & In-space            & Fully kinetic  & $4.0\times 10^{14}$ \si{m^{-3}} & $5.63 \times 10^8$ s$^{-1}$ \\
1A      & In-chamber          & Fully kinetic  & $4.0\times 10^{13}$ \si{m^{-3}} & $1.78 \times 10^8$ s$^{-1}$ \\
2A      & In-chamber          & Fully kinetic  & $4.0\times 10^{14}$ \si{m^{-3}} & $5.63 \times 10^8$ s$^{-1}$ \\
1B      & In-chamber          & Boltzmann      & $4.0\times 10^{13}$ \si{m^{-3}} & $1.78 \times 10^8$ s$^{-1}$ \\
2B      & In-chamber          & Boltzmann      & $4.0\times 10^{14}$ \si{m^{-3}} & $5.63 \times 10^8$ s$^{-1}$ \\
\hline \hline
\end{tabular}
\label{tab:caseid}
\end{table}

Table~\ref{tab:thrustercondition} shows the thruster exit condition of each species type. The number density ratio of neutrals to ions is equal to 20, and the ratio of ions to electrons is equal to four, similar to our previous works~\cite{Jambunathan2020-io, Nuwal2020-pa}. At this inflow BC, the neutral particles and electrons are assumed to have a uniform profile in density and initialized with a full Maxwellian in the cross-stream direction and a half-Maxwellian in the streamwise direction. The ions have a 12$^\circ$ Gaussian distribution in number density versus radial distance from the center of the thruster~\cite{Hyakutake2003-oc}, and the thermal distribution of the ion beam at the thruster exit is neglected. In the high-density conditions (0A, 2A, and 2B), the neutral particle flux at the thruster exit is $j_\mathrm{n}=1.6\times 10^{18}$ \si{m^{-2} s^{-1}}, whereas the average ion particle flux at the thruster exit is  $j_\mathrm{ex}=1.0\times 10^{19}$ \si{m^{-2} s^{-1}}. According to Eq~\eqref{eq:propeff}, $\eta_\mathrm{u} = j_\mathrm{ex} / \left(j_\mathrm{ex} + j_\mathrm{n}\right)=0.86$, which is similar to that of well-established thrusters~\cite{Polk2000-dk,Hickman2005-ha}. For the Boltzmann cases, the reference potential ($\phi_0$ in Eq.~\eqref{eq:boltz}) is set at 0 V since the exit plasma potential must be the same as that of the thruster grid if the beam is completely neutralized. Similarly, $T_\mathrm{e0}$ is 23,200 K (2 eV), and $n_\mathrm{0}$ is the same as maximum ion number density shown in Table~\ref{tab:caseid}.

\begin{table}[hbt!]
\caption{Thruster exit parameters for each speciess.}
\centering
\begin{tabular}{cccc}
\hline \hline
Thruster exit conditions & Xe & Xe$^+$ & e$^-$$^*$ \\ \hline
Bulk velocity (m/s) & 200 & 40,000 & 0 \\
Density, $n_\mathrm{ex}/n_0$$^\dagger$ & 20 & 1.0 & 0.25 \\
Temperature (K) & 300 & - & 23,210 (2 eV)\\
Inlet type & Half-Maxwellian & 12$^\circ$ Gaussian beam & Half-Maxwellian \\
\hline \hline
\multicolumn{4}{l}{$^*$ Electrons are only modeled in the fully kinetic cases (0A, 1A, and 2A).} \\
\multicolumn{4}{l}{$^\dagger$ $n_0$ is $4.0\times 10^{13}$ m$^{-3}$ for the 1A and 1B cases, and $4.0\times 10^{14}$ m$^{-3}$ for the 0A, 2A, and 2B cases.} \\
\end{tabular}
\label{tab:thrustercondition}
\end{table}

Table~\ref{tab:computationalparameters} shows the PIC-DSMC parameters for all cases. In the kinetic simulations, the ion and electron timesteps should follow $\Delta t_e < 0.1/\omega_\mathrm{pe}$. $\omega_\mathrm{pe}$ is plasma frequency calculated as $\omega_\mathrm{pe}=\sqrt{n_0e^2/m_\mathrm{e}\varepsilon_0}$, where $e$, $m_\mathrm{e}$, and $\varepsilon_0$ are the elementary charge, the mass of an electron, and the permittivity of the vacuum. The plasma frequency for each case is shown in Table~\ref{tab:caseid}. In this work, we choose a timestep of $2.8 \times 10^{-10}$ s for the low-density case (1A) and $1.0 \times 10^{-10}$ s for the high-density cases (0A and 2A), similar to our previous work~\cite{Nuwal2020-pa}. In the time-slicing method of the DSMC and PIC modules, described in Section~\ref{subsec:pic-dsmccounple}, $F_\mathrm{col}=100$ is used for the fully kinetic simulations (0A, 1A, and 2A). In this study, the criterion for the E-FOT cell size is $\Delta x < \lambda_D$ for a first-order accurate PIC result, resulting in the minimum cell size of $1.5\times 10^{-3}$ m for the low-density case (1A) and $7.8\times 10^{-4}$ m for the high-density cases (0A and 2A). For the Boltzmann cases, we use timesteps of $2.8\times 10^{-8}$ s regardless of the number density and larger cell size as the criterion for the E-FOT since electrons are not explicitly modeled under the quasi-neutral assumption. The superparticle factor, $F_\mathrm{num}$, and species weight, $W$, are chosen such that there are at least 15 particles per species per cell both for C-FOT and E-FOT, as shown in Table~\ref{tab:computationalparameters}. 

\begin{table}[hbt!]
\caption{PIC-DSMC parameters for all cases.}
\centering
\begin{tabular}{cccccc}
\hline \hline
Simulation parameters & 0A & 1A & 2A & 1B & 2B \\ \hline
Xe timestep, $\Delta t_\mathrm{n}$ (s) & $4.0\times 10^{-6}$ & $5.6\times 10^{-6}$ & $4.0\times 10^{-5}$ & $5.6\times 10^{-6}$ & $5.6\times 10^{-6}$ \\
Xe$^+$/e$^-$ timestep, $\Delta t_\mathrm{i}=\Delta t_\mathrm{e}$ (s) & $1.0\times 10^{-10}$  & $2.8\times 10^{-10}$ & $1.0\times 10^{-10}$ & $2.8\times 10^{-8}$ & $2.8\times 10^{-8}$ \\
DSMC interval, $F_\mathrm{col}$ & 100 & 100 & 100 & 1 & 1 \\
Superparticle factor, $F_\mathrm{num}$ & $1.2\times 10^6$ & $1.0\times 10^6$ & $1.2\times 10^7$ & $1.0\times 10^6$ & $2.0\times 10^7$ \\
Species weight, $W_i/W_n$$^*$ & 0.0025 & 0.005 & 0.00025 & 0.005 & 0.005 \\
\hline \hline
\multicolumn{6}{l}{$^*$ $W_n=1$ in all cases.} \\
\end{tabular}
\label{tab:computationalparameters}
\end{table}

CHAOS uses multiple GPUs with MPI-Cuda parallelization strategies. The demonstration of strong scaling studies is given in Ref.~\cite{Jambunathan2018-xf}. This study used NVIDIA A100 GPUs on Delta at the National Center for Supercomputing Applications for all cases. For the 0A and 2A cases, higher-density kinetic cases, the total simulation runtimes to obtain a steady PIC field were about 40 hours with 64 GPUs. An additional calculation to obtain a steady DSMC field took about two days. For the 1A, 1B, and 2B cases, we used 16 GPUs, and the total simulation runtimes were about two days, including steady-field sampling. The sputtering calculation cost was reduced by restarting the calculation from the steady-state DSMC results with a total sputtering simulation runtime of about 5 hours.


\section{Results and Discussions}
\label{sec:result}

\subsection{Effect of Ground Facility Boundary Condition on the Ion Plumes}
\label{sec:result-spacechamber}

\begin{figure}[hbt!]
    \begin{subfigure}{0.5\textwidth}
        \centering
        \includegraphics[width=\linewidth]{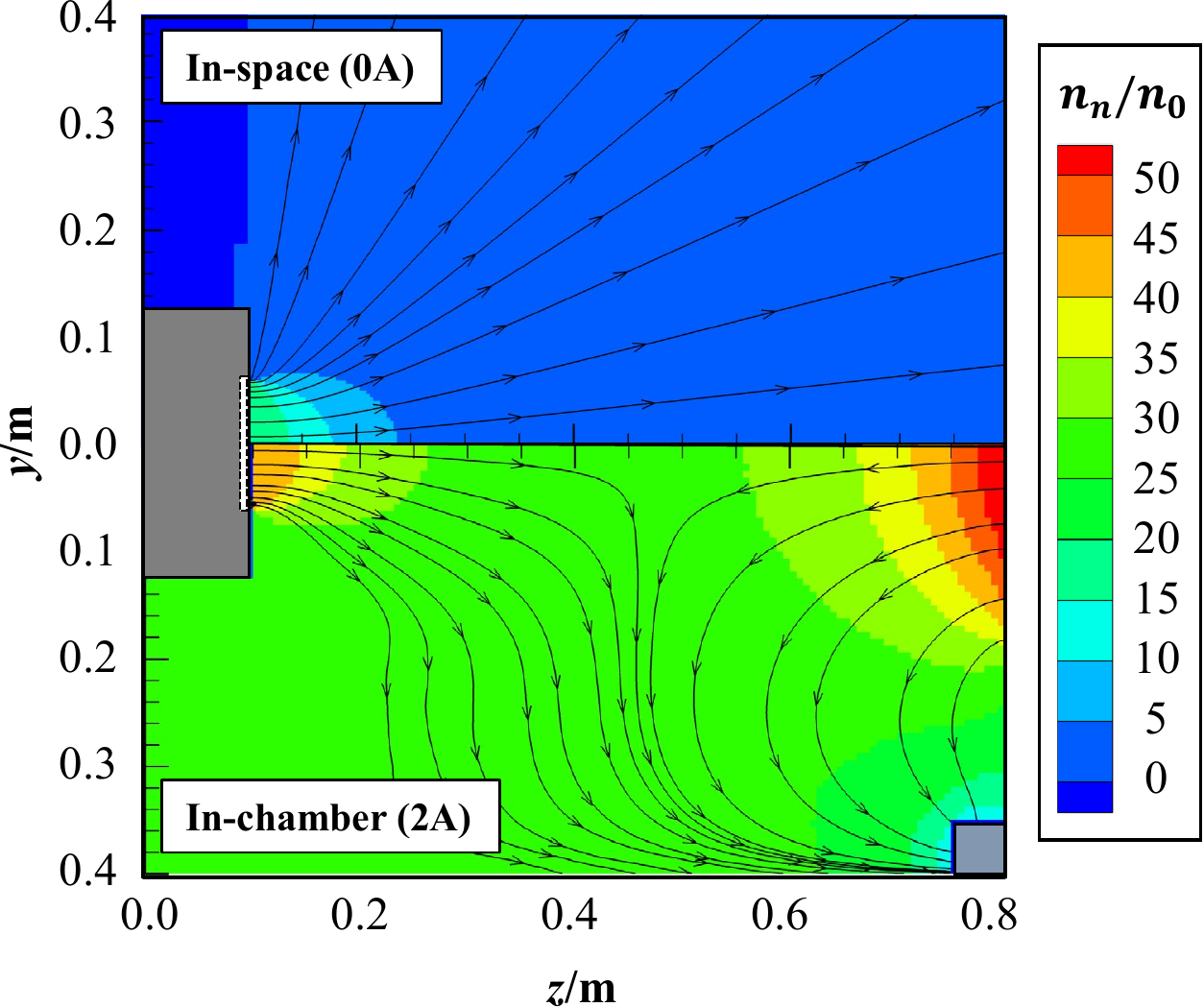}
        \caption{In the thruster center plane ($x=0$ m)} 
        \label{fig:nd_2d}
    \end{subfigure}%
    \begin{subfigure}{0.5\textwidth}
        \centering
        \includegraphics[width=0.95\linewidth]{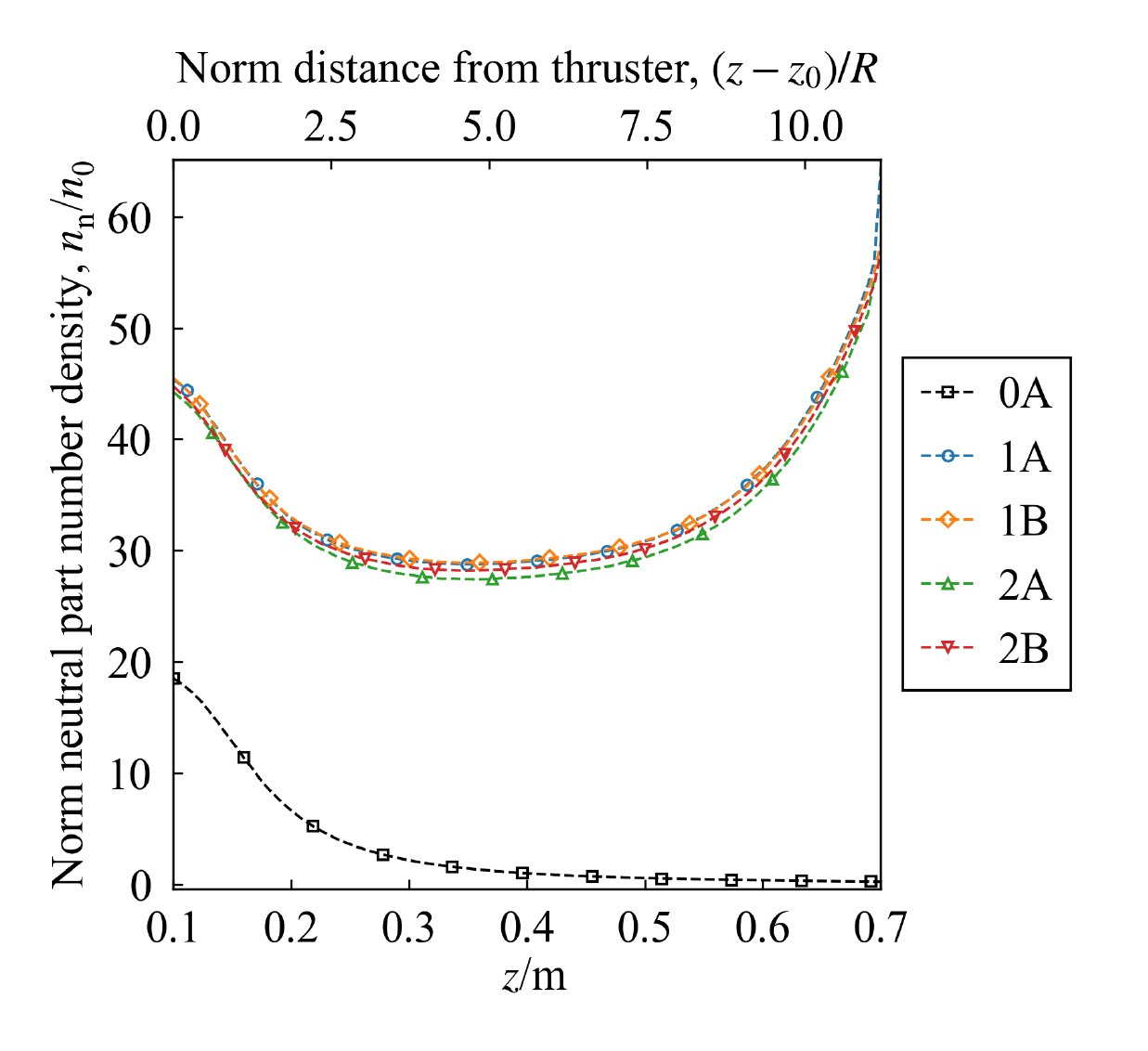}
        \caption{Along the thruster axis ($x=0$ m, $y=0$ m)} 
        \label{fig:nd_centerline}
    \end{subfigure}%
    \caption{Neutral particle (Xe) number density comparison between the in-space (0A) and in-chamber cases (1A, 2A, 1B, and 2B), where $\mathbf{\mathit{n}_0 = 4.0\times 10^{13}}$ m$^{-3}$ in the 1A and 1B cases, and $\mathbf{\mathit{n}_0 = 4.0\times 10^{14}}$ m$^{-3}$ in the 0A, 2A, and 2B cases. The streamlines shown in (a) are obtained from the neutral particle velocities. $z_0$ and $R$ in (b) are the thruster exit $z$-position of 0.1 m and the thruster radius of 0.0625 m, respectively} 
    \label{fig:neutraldensity}
\end{figure}

This section discusses how the ground facility environment affects the ion thruster plume compared to the space environment. The first large difference is the neutral background density. Unlike in-space operation, heavy particles stay inside the vacuum chamber until exhausted through the vacuum pumps, resulting in finite background pressure. Figure~\ref{fig:nd_2d} shows the normalized Xe neutral particle number density in the $x=0$ m plane for the in-space (0A) versus in-chamber (2A) cases. Although the thruster exit neutral density is set at $n_\mathrm{n}/n_0=20$, the neutral density of the in-chamber case is higher than that of the in-space case. The streamlines in Fig.~\ref{fig:nd_2d} show the $z$- and $y$-velocity of neutral particles. In the in-chamber case, the ejected neutral particles flow to the bottom-right numerical pump, in contrast to the in-space case. At the downstream wall, neutralized ions create another neutral particle source. The normalized number density on the thruster axis is shown in Fig.~\ref{fig:nd_centerline} as a function of the normalized distance from the thruster exit. As seen from this plot, the background neutral density increases by a factor of 10--100 or more in the ground test, and its distribution also changes. The figure also includes the results of other in-chamber cases, 1A, 1B, and 2B. There is little difference in the background neutral density distribution normalized by the exit ion density, regardless of the density and the method used to calculate the potential. The average background pressure in the chamber is 0.034 \si{\micro Torr} for the 1A and 1B cases and 0.34 \si{\micro Torr} for the 2A and 2B cases.

\begin{figure}[hbt!]
    \begin{subfigure}{0.5\textwidth}
        \centering
        \includegraphics[width=\linewidth]{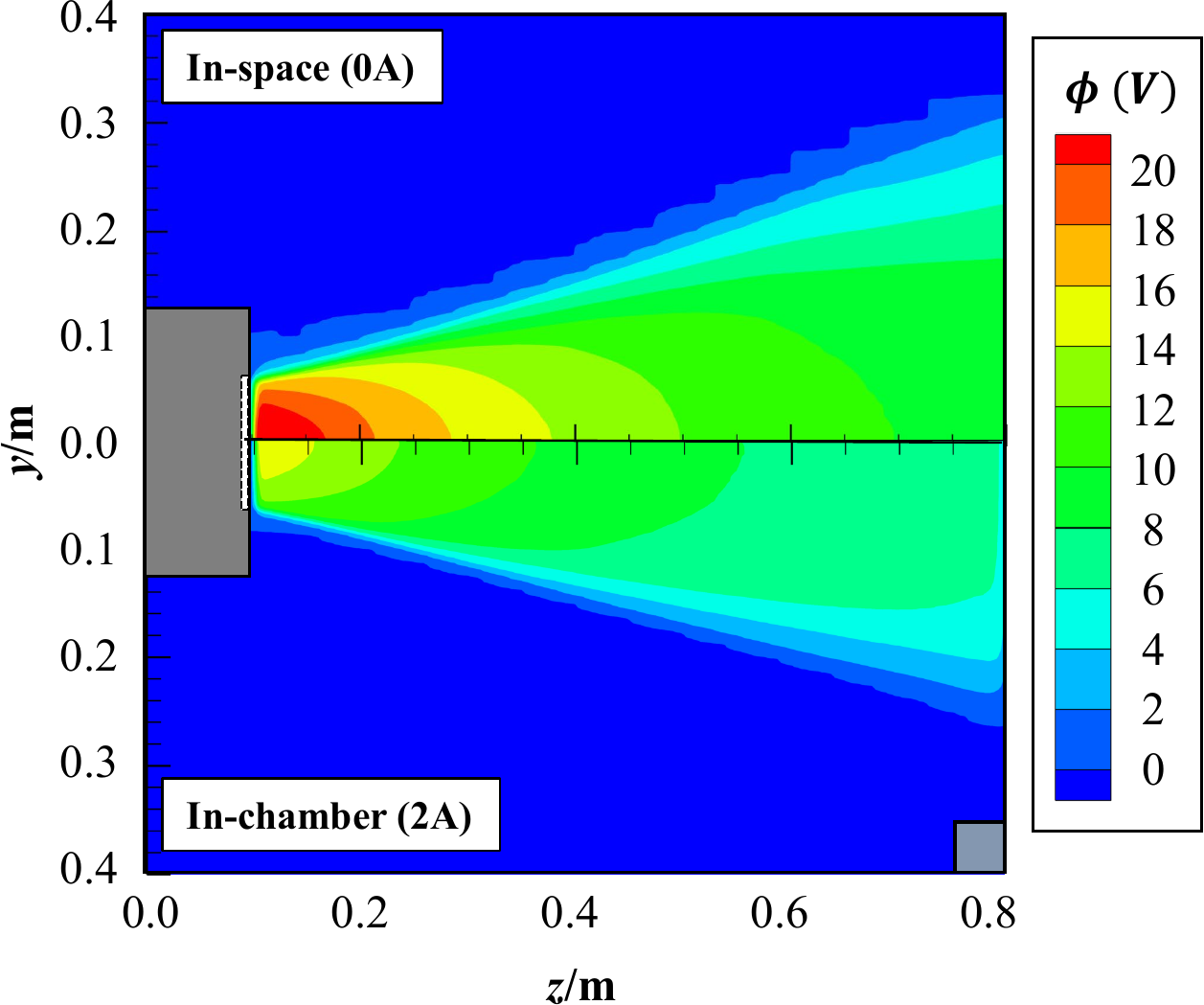}
        \caption{In the thruster center plane ($x=0$ m)} 
        \label{fig:phi0A2A_2d}
    \end{subfigure}%
    \begin{subfigure}{0.5\textwidth}
        \centering
        \includegraphics[width=0.9\linewidth]{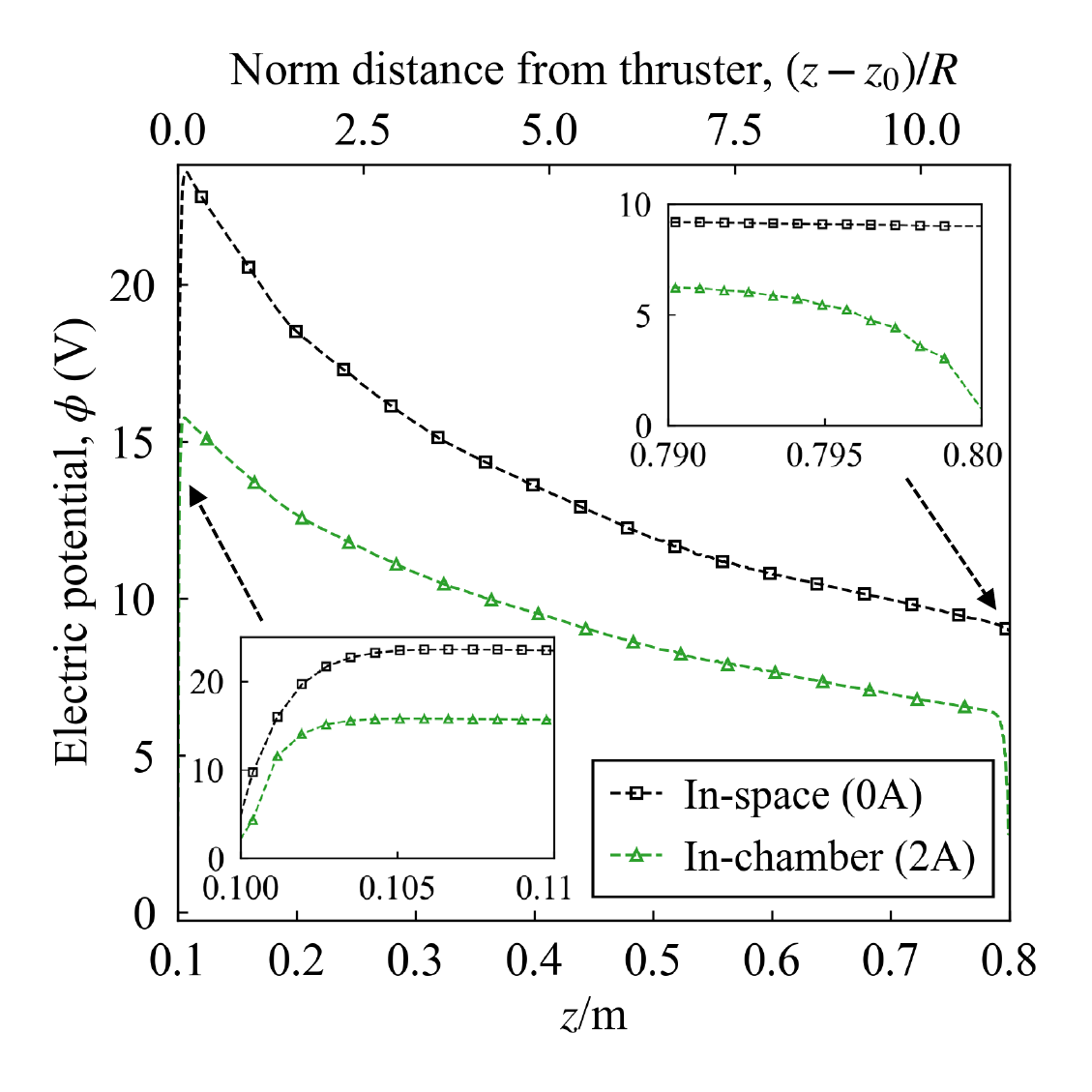}
        \caption{Along the thruster axis ($x=0$ m, $y=0$ m)} 
        \label{fig:phi0A2A_centerline}
    \end{subfigure}%
    \caption{Electric potential comparison between the in-space (0A) and in-chamber (2A) cases. $z_0$ and $R$ in (b) are the thruster exit $z$-position of 0.1 m and the thruster radius of 0.0625 m, respectively.} 
    \label{fig:electricpotential_0A2A}
\end{figure}

The grounded vacuum chamber walls also affect the electric potential of the plasma plume, which changes the flow of electrons with small mass and CEX ions with small velocity. Figure~\ref{fig:phi0A2A_2d} shows the electric potential contours in the $x=0$ m plane for the in-space (0A) versus in-chamber (2A) cases. The in-space case yields a higher potential in the plume core region due to differences in the downstream electric BC. Figure~\ref{fig:phi0A2A_centerline} shows the line plots of the electric potential on the thruster axis. The neutralization region near the thruster exit ($(z-z_0)/R < 0.05$), where the ion density is larger than the electron density, shows the formation of a sheath and an increase in electric potential (see the enlarged plot at the lower left in Fig.~\ref{fig:phi0A2A_centerline}). Looking at the region close to the downstream wall ($(z-z_0)/R > 11$), a sheath is also formed in the chamber due to the Dirichlet and charge-absorbing particle BCs on the edge of the domain, in contrast to the in-space case with the Neumann electrical BC (see the enlarged plot at the upper right in Fig.~\ref{fig:phi0A2A_centerline}). The downstream wall with a lower potential than the plume decreases the electric potential in the entire plume.

\begin{figure}[hbt!]
    \begin{subfigure}{0.55\textwidth}
        \centering
        \includegraphics[width=\linewidth]{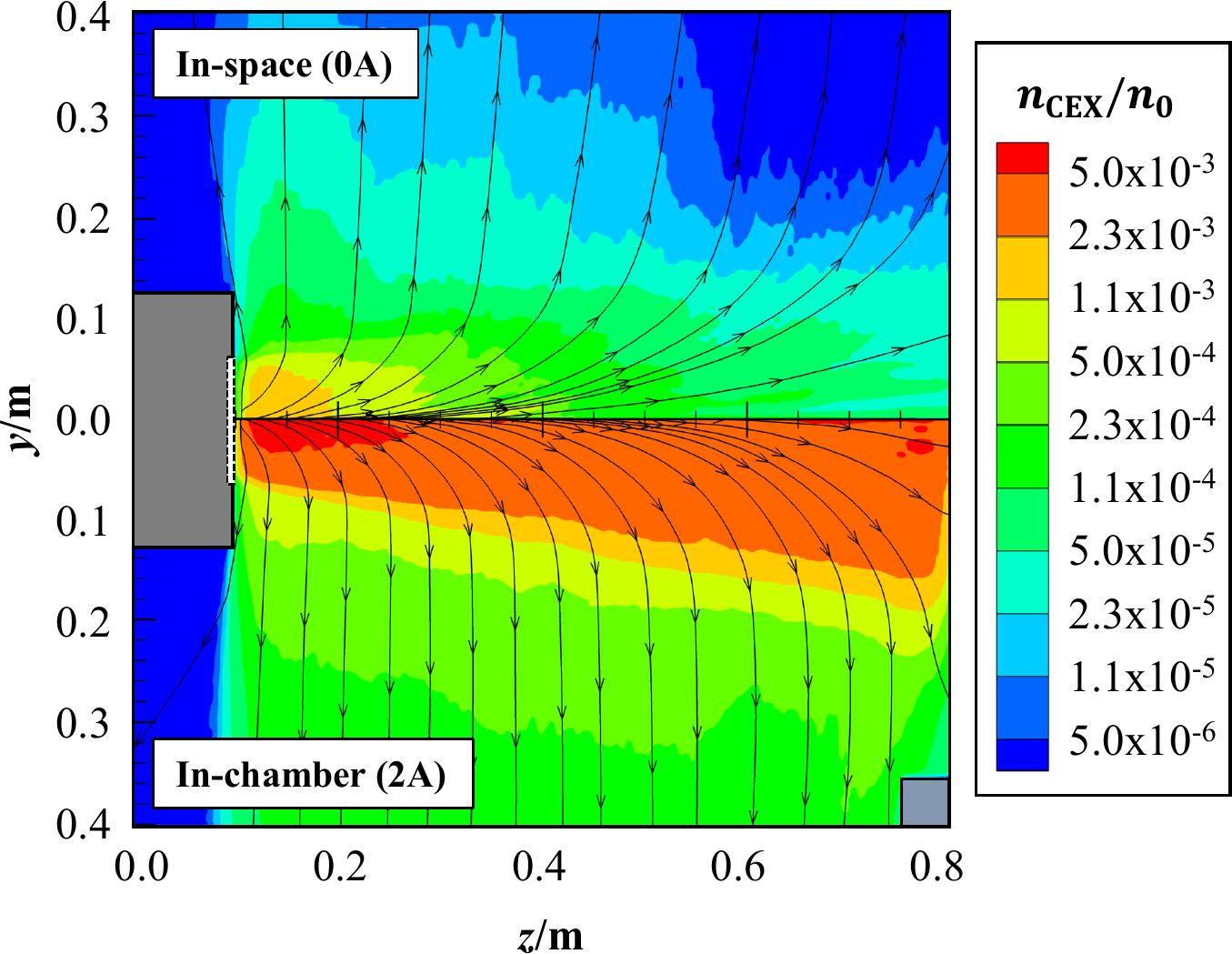}
        \caption{In the thruster center plane ($x=0$ m)} 
        \label{fig:nCEX_0A2A_2d}
    \end{subfigure}%
    \begin{subfigure}{0.45\textwidth}
        \centering
        \includegraphics[width=\linewidth]{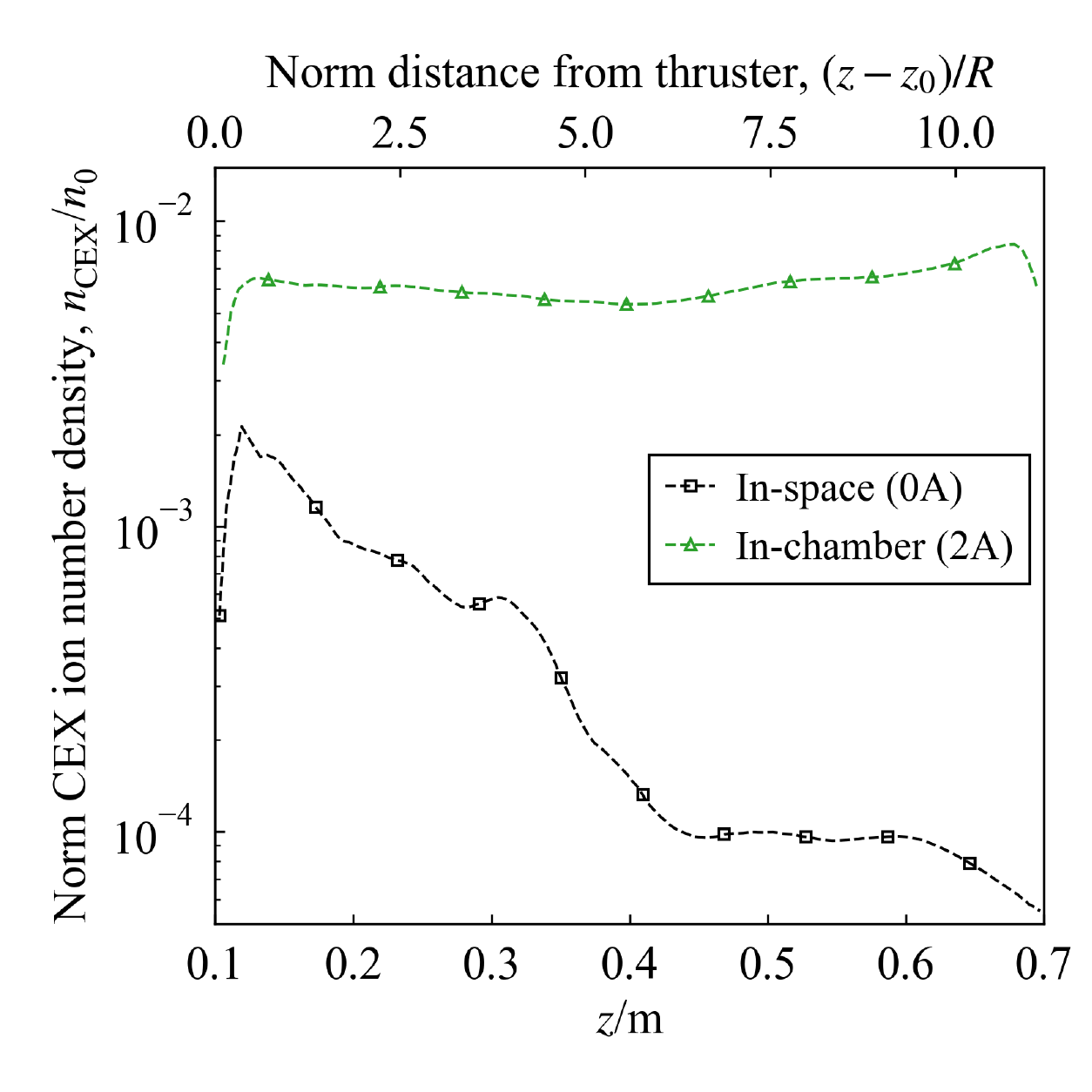}
        \caption{Along the thruster axis ($x=0$ m, $y=0$ m)} 
        \label{fig:nCEX_0A2A_line}
    \end{subfigure}%
    \caption{CEX ion number density comparison between the in-space (0A) and in-chamber (2A) cases, where $\mathbf{\mathit{n}_0 = 4.0\times 10^{14}}$ m$^{-3}$. The streamlines shown in (a) are obtained from the CEX ion velocities. $z_0$ and $R$ in (b) are the thruster exit $z$-position of 0.1 m and the thruster radius of 0.0625 m, respectively.} 
    \label{fig:CEXdensity_0A2A}
\end{figure}

The ground facility effects shown in this section change the population of CEX ions, one of the major causes of spacecraft backflow contamination. Figure~\ref{fig:nCEX_0A2A_2d} shows the normalized number density and streamlines of CEX ions in the $x=0$ m plane for the in-space (0A) versus in-chamber (2A) cases. In the in-space case, many CEX ions are produced close to the thruster and spread to the downstream and off-axis region. In the in-chamber case, however, a large density region exists downstream in the ion beam, and the CEX ions are more spread in the cross-stream direction. The CEX ion number density in the in-chamber case is only a few times larger near the thruster exit but about 100 times larger than the in-space case at $(z-z_0)/R \sim 10$, as shown in Fig.~\ref{fig:nCEX_0A2A_line}. This is because the neutral background density in the chamber is large downstream, as shown in Fig.~\ref{fig:neutraldensity}, resulting in more of the CEX ions being produced downstream. The CEX ion density decreases near the thruster ($(z-z_0)/R < 0.5$), and the rate of decrease is larger in space. This suggests that the larger potential gradient created by the sheath in space makes more CEX ions flow back onto the thruster, as shown in the enlarged plot at the lower left in Fig.~\ref{fig:phi0A2A_centerline}.

\subsection{Effect of Boltzmann Assumption on the Ion Plumes in the Ground Facility}
\label{sec:result-kineticBoltz}

\begin{figure}[hbt!]
    \begin{subfigure}{0.5\textwidth}
        \centering
        \includegraphics[width=\linewidth]{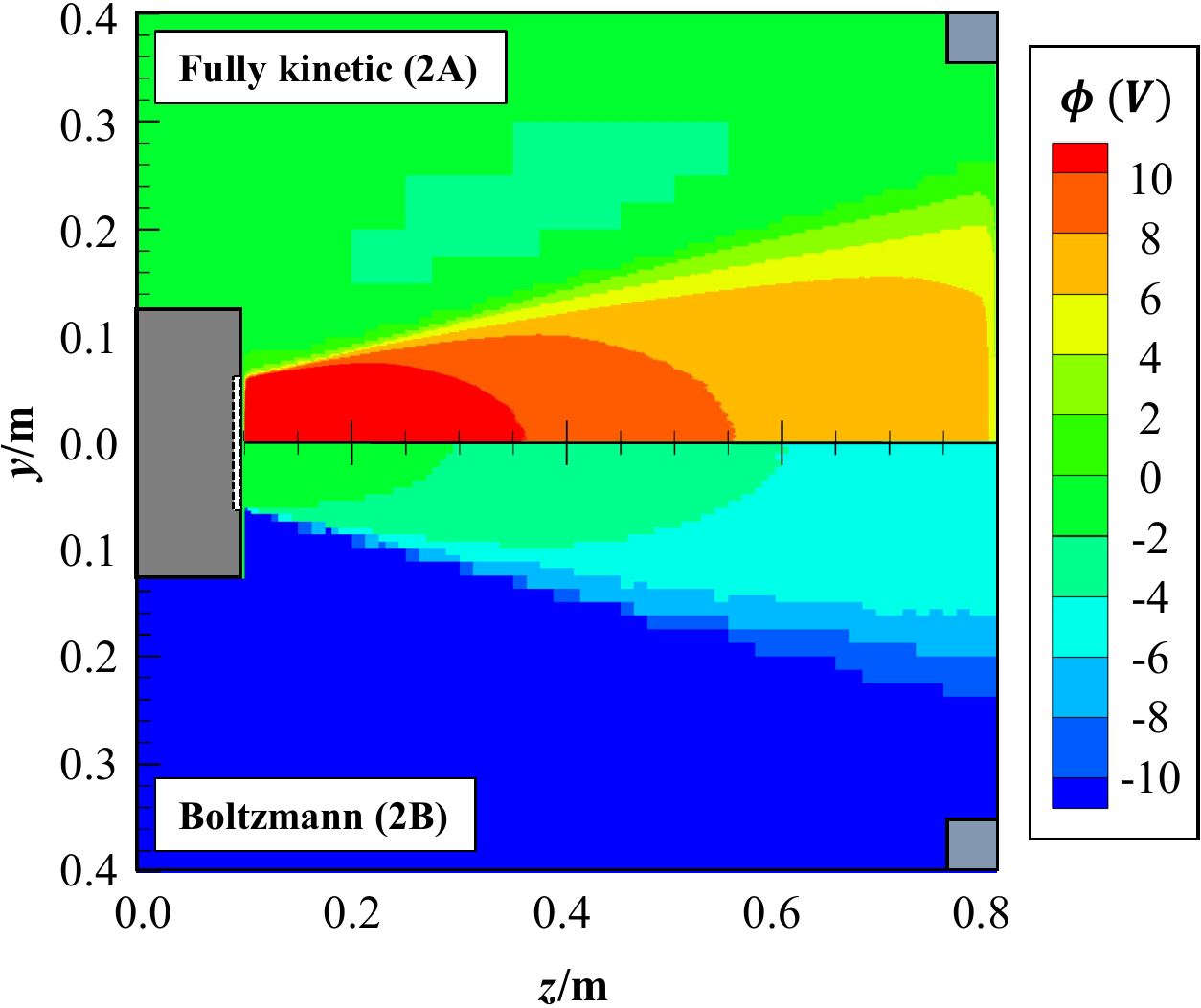}
        \caption{In the thruster center plane ($x=0$ m)} 
        \label{fig:phi2A2B_2d}
    \end{subfigure}%
    \begin{subfigure}{0.5\textwidth}
        \centering
        \includegraphics[width=0.87\linewidth]{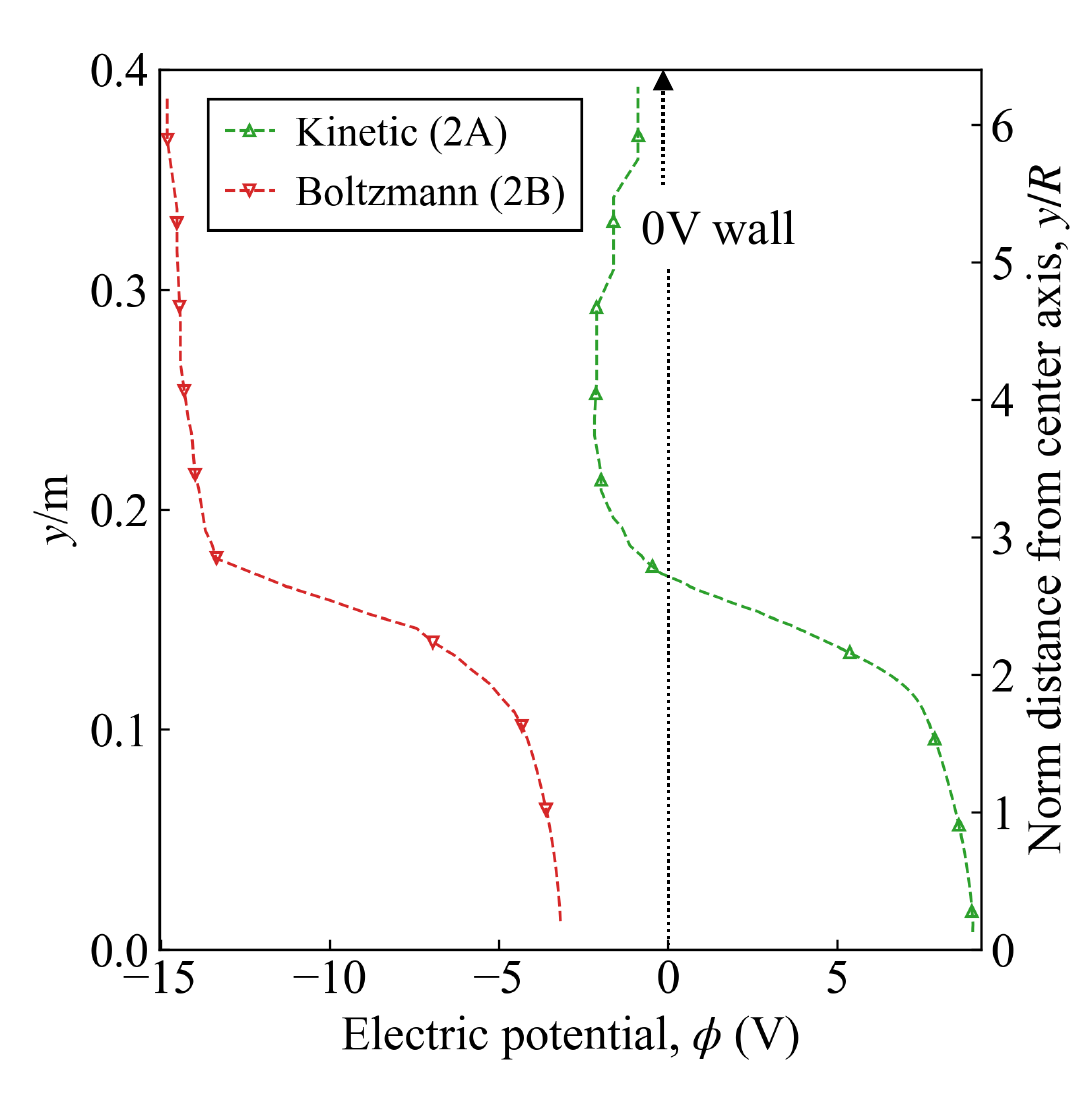}
        \caption{Along the $y$-direction ($x=0$ m, $z=0.45$ m)} 
        \label{fig:phi2A2B_line}
    \end{subfigure}%
    \caption{Electric potential comparison between the fully kinetic (2A) and Boltzmann (2B) cases. $R$ in (b) is the thruster radius of 0.0625 m.} 
    \label{fig:electricpotential_2A2B}
\end{figure}

This section compares the results obtained from a fully kinetic simulation to that obtained from the quasi-neutral Boltzmann simulation, which has been commonly used to simulate the electric propulsion thruster plumes. Figure~\ref{fig:phi2A2B_2d} shows the electric potential contours in the $x=0$ m plane for the fully kinetic (2A) versus Boltzmann (2B) cases. It can be seen that the maximum potential is 15 V near the thruster exit and approximately 0 V outside the plume in the fully kinetic case. In the Boltzmann case, however, the maximum potential is 0 V near the thruster exit and approximately -15 V outside the plume. This difference is because the quasi-neutral Boltzmann case does not reproduce the sheath between the plume and thruster exit. The electric potential in the cross-stream direction ($y$-direction in Fig.~\ref{fig:phi2A2B_2d}) at $z=0.45$ m is shown in Fig.~\ref{fig:phi2A2B_line}. While the potential difference between the center of the plume and its edge is about -10 V in both cases, the minimum and maximum potentials differ significantly.

\begin{figure}[hbt!]
    \begin{subfigure}{0.54\textwidth}
        \centering
        \includegraphics[width=\linewidth]{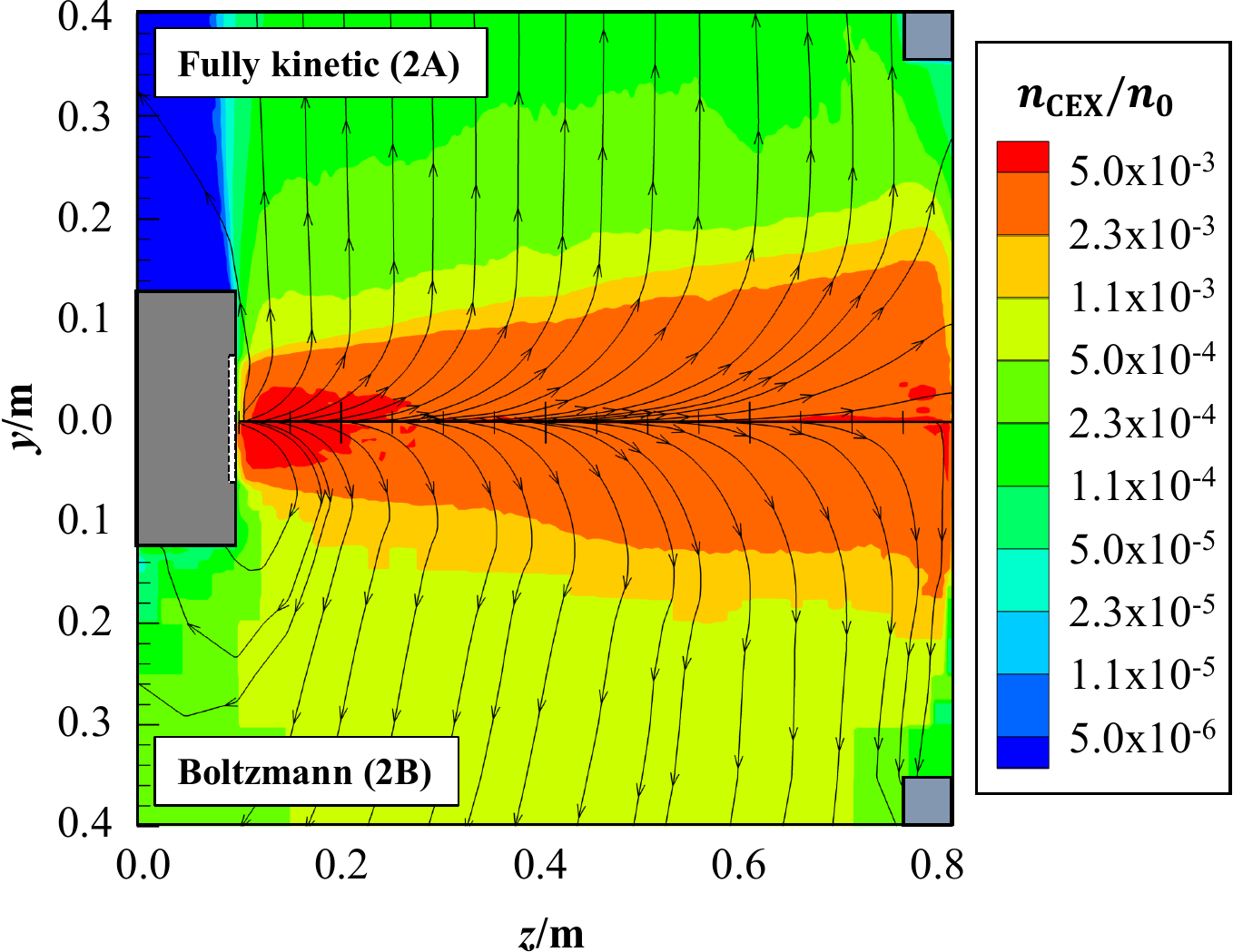}
        \caption{In the thruster center plane ($x=0$ m)} 
        \label{fig:nCEX_2A2B_2d}
    \end{subfigure}%
    \begin{subfigure}{0.46\textwidth}
        \centering
        \includegraphics[width=0.95\linewidth]{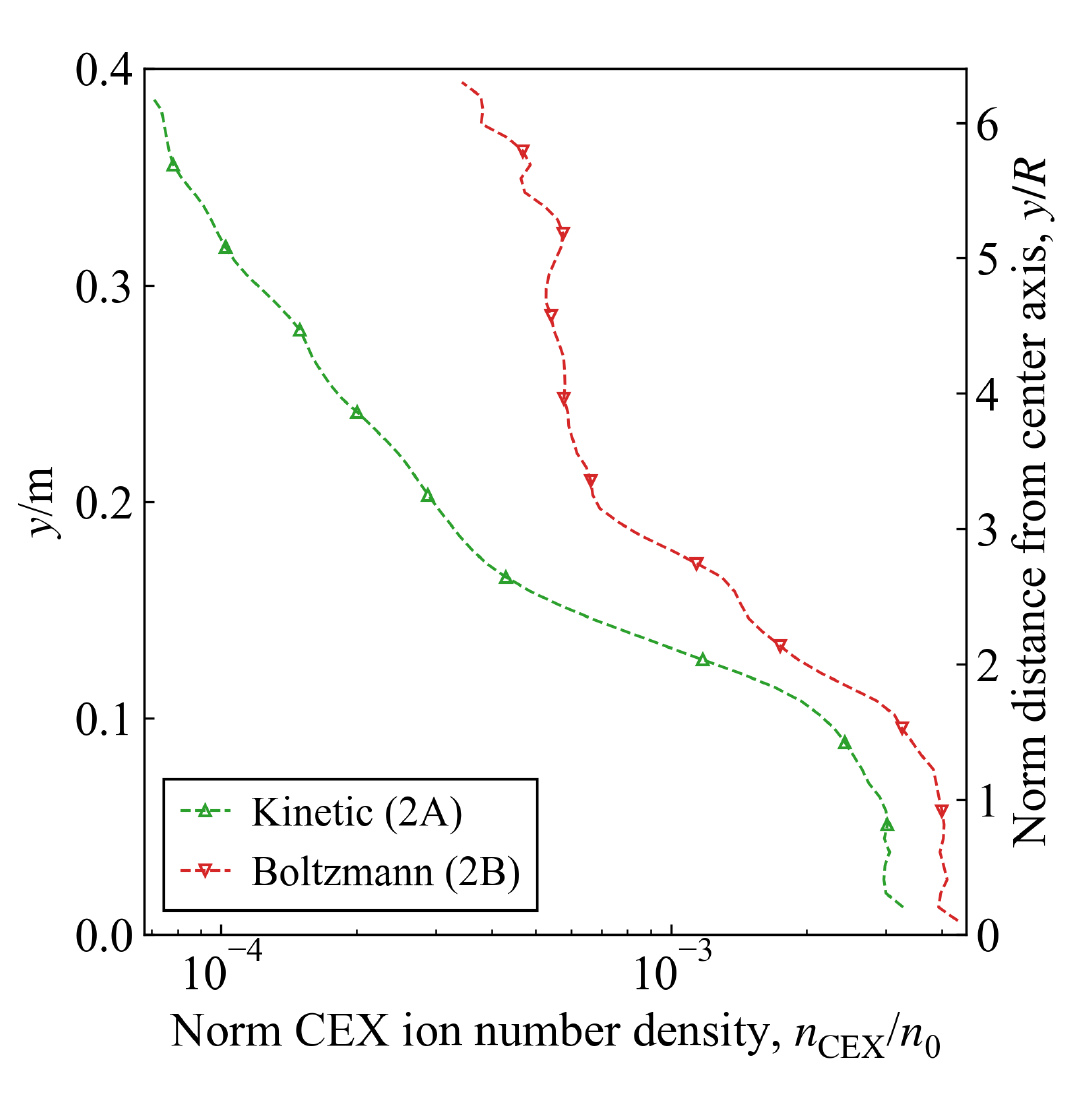}
        \caption{Along the $y$-direction ($x=0$ m, $z=0.45$ m)} 
        \label{fig:nCEX_2A2B_line}
    \end{subfigure}%
    \caption{CEX ion number density comparison between the fully kinetic (2A) and Boltzmann (2B) cases, where $\mathbf{\mathit{n}_0 = 4.0\times 10^{14}}$ m$^{-3}$. The streamlines shown in (a) are obtained from the CEX ion velocities. $R$ in (b) is the thruster radius of 0.0625 m.} 
    \label{fig:CEXdensity_2A2B}
\end{figure}

Next, we investigate how this difference in electric potential affects the facility plasma, i.e., the ions produced by collisions between ions and neutral particles. Figure~\ref{fig:nCEX_2A2B_2d} shows the normalized number density and streamlines of CEX ion on the $x=0$ m plane for the fully kinetic (2A) and Boltzmann (2B) cases, respectively. The streamlines of CEX ions are mostly perpendicular to the $y$-max wall in the fully kinetic case, whereas in the Boltzmann case, the streamlines seem to go behind the thruster. This is because the potential becomes small in the region of low ion density using the Boltzmann relation (Eq.~\eqref{eq:boltz}), attracting the slow CEX ions into the small potential region. Looking at the CEX ion number density outside the beam ion plume, the CEX ion density is higher in the Boltzmann case. Figure~\ref{fig:nCEX_2A2B_line} shows the CEX ion number density in the cross-stream direction ($y$-direction on the Fig.~\ref{fig:nCEX_2A2B_2d} plane) at $z=0.45$ m. A comparison of the fully kinetic and the Boltzmann cases shows that the difference in the number density is about 1.3 times larger inside the beam ion plume but 5 times larger outside the plume for the Boltzmann case. The potential difference shown in Fig.\ref{fig:electricpotential_2A2B} explains this density difference. Since the outer walls of the chamber are subject to a Dirichlet condition of 0 V, a negative electric field is formed near the wall in the Boltzmann case, which traps slow CEX ions in the vacuum chamber.

\begin{figure}[hbt!]
    \centering
    \includegraphics[width=\textwidth]{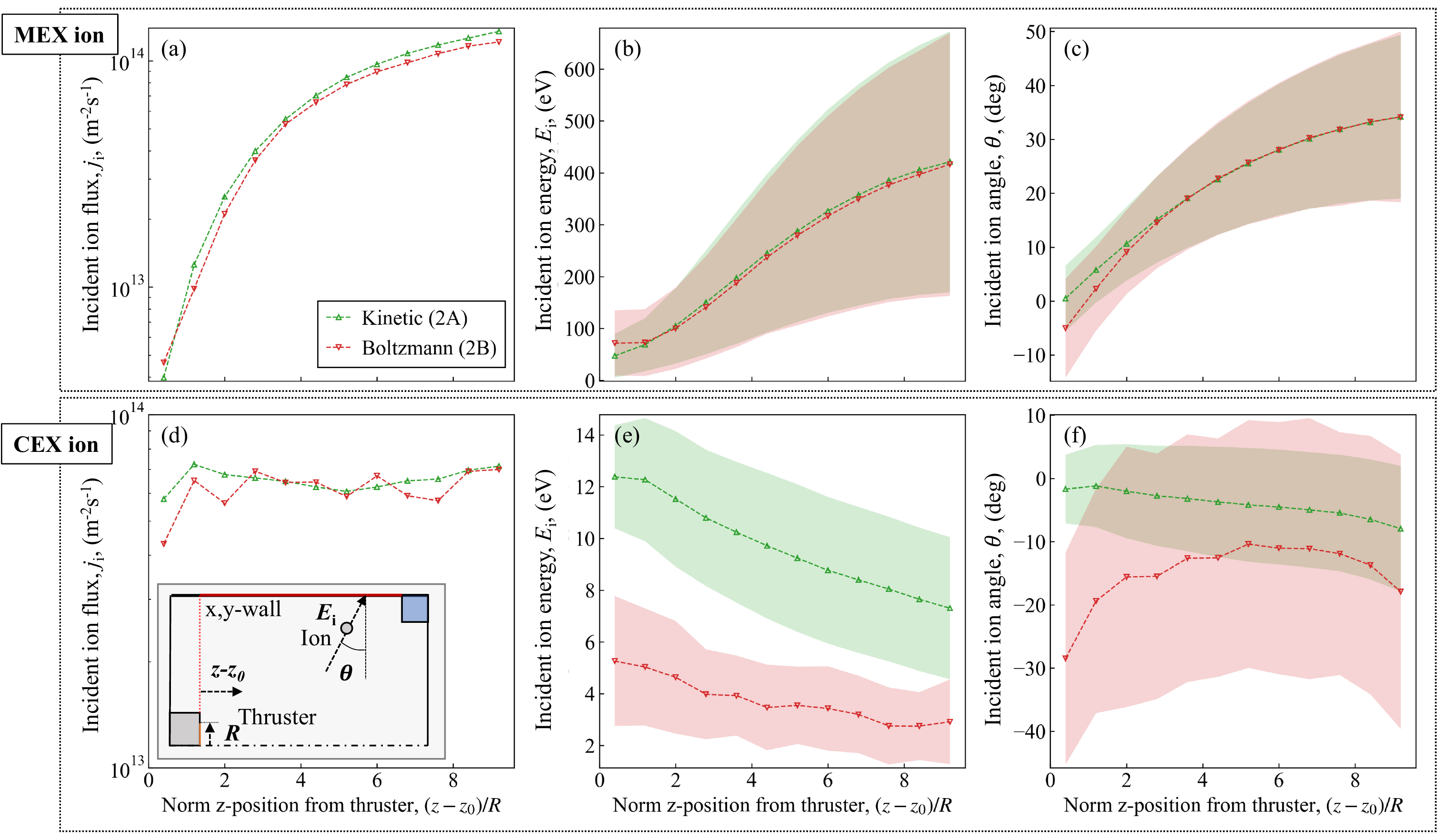}
    \caption{Particle flux, energy, and angle of the incident ions on the side walls ($x$ = 0.4 m and $y$ = 0.4 m) for the MEX ions (top) and CEX ions (bottom) in the 2A and 2B cases. On the energy and angle plots, the markers indicate the average values, and the color bands indicate the standard deviation in each bin. $z_0$ and $R$ are the thruster exit $z$-position of 0.1 m and the thruster radius of 0.0625 m, respectively.}
    \label{fig:sidewallincident_2A2B}
\end{figure}

Since we are interested in sputtering, we directly sample the computational ion particles to investigate the particle flux, energy, and angle of the facility plasma incident on the side walls at $x=0.4$ m and $y=0.4$ m. Due to the large velocity of 40,000 m/s in the streamwise direction, beam ions do not reach the side walls unless they collide with neutral particles. Figure~\ref{fig:sidewallincident_2A2B} shows the flux, energy, and angle of incident ions on the $x$ = 0.4 m and $y$ = 0.4 m walls for MEX ions (top) and CEX ions (bottom). For the incident energy and angle, the markers and color bands indicate the average values and the standard deviations, respectively. The detailed energy distribution and angle distribution in each bin are given in the appendix section~\ref{subsec:sidehitEDF} for the 2A case. The particle fluxes of these two types of ions are both on the order of $10^{13}$--$10^{14}$ /m$^2$s, which is very small compared to the flux of the beam ions ($j_\mathrm{ex}=1.0 \times 10^{19}$ /m$^2$s). As shown in Figs~\ref{fig:sidewallincident_2A2B}a-c, for the MEX ions, there is almost no difference between the fully kinetic and Boltzmann cases. Because MEX ions originally have relatively large energy, the electric potential difference has a small impact on the trend of MEX results. The incident flux of MEX ions increases as the distance increases because ions with larger $z$-velocities travel long distances in the $z$-direction before hitting walls. The incident energies and angles tend to increase similarly for the same reason as the flux. In contrast to the MEX ions, CEX ions, initially slow neutral particles, acquire energy mostly from the electric potential and move along the electric field shown in Fig~\ref{fig:nCEX_2A2B_2d}. Since the amount of CEX ions produced is almost the same in the fully kinetic and Boltzmann cases, the fluxes are comparable, but the incident energies and angles differ, as shown in Figs.~\ref{fig:sidewallincident_2A2B}d-f. As we explain for Fig~\ref{fig:electricpotential_2A2B}, in the Boltzmann case, the potential barrier on the chamber reduces the incident energy. The incident angle is reduced at $(z-z_0)/R<2$ due to the CEX ions going behind the thruster, where the ion density is small.

In this study, the reference potential, $\phi_0$, is set to 0 V, the voltage at the thruster exit. However, as seen in the kinetic simulation, the maximum potential is around 15 V. If we use this value as $\phi_0$, the potential distribution in the Boltzmann case would be closer to that obtained by the kinetic simulation, except for at the sheath region. Nevertheless, obtaining this reference potential requires prior kinetic simulations or experiments, and it is impossible to perform self-consistent calculations using a Boltzmann simulation alone. Therefore, the fully kinetic simulation remains a crucial tool to capture the potential of EP plumes accurately.


\subsection{Effect of Thruster Exit Number Density on the Ion Plumes in the Ground Facility}
\label{sec:result-numdens}

\begin{figure}[hbt!]
    \begin{subfigure}{0.5\textwidth}
        \centering
        \includegraphics[width=\linewidth]{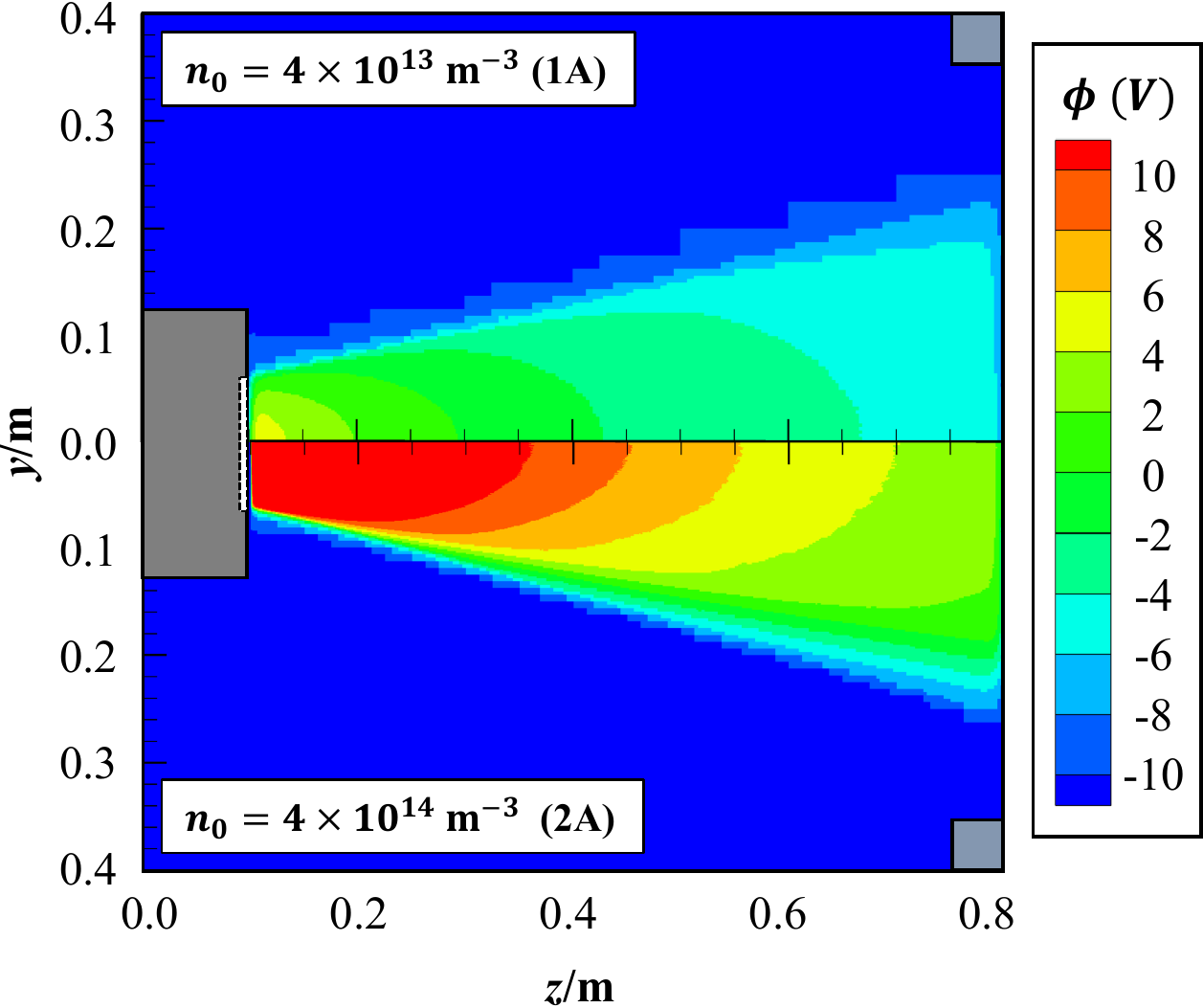}
        \caption{In the thruster center plane ($x=0$ m)} 
        \label{fig:phi1A2A_2d}
    \end{subfigure}%
    \begin{subfigure}{0.5\textwidth}
        \centering
        \includegraphics[width=0.87\linewidth]{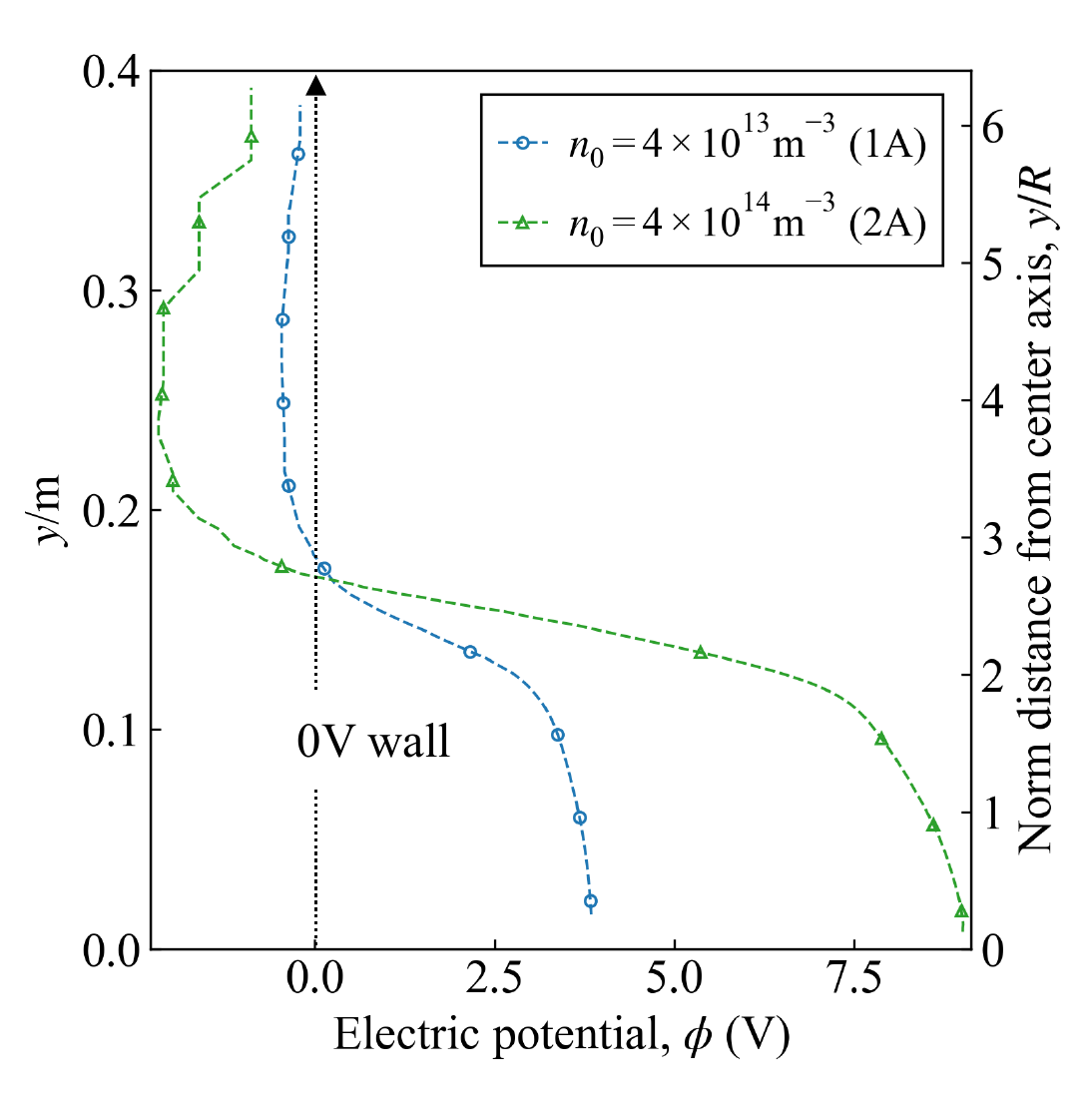}
        \caption{Along the $y$-direction ($x=0$ m, $z=0.45$ m)} 
        \label{fig:phi1A2A_line}
    \end{subfigure}%
    \caption{Electric potential comparison between the low-density (1A) and high-density (2A) cases. $R$ in (b) is the thruster radius of 0.0625 m.}
    \label{fig:electricpotential_1A2A}
\end{figure}

This section investigates the effect of thruster exit number density on the facility plasma using a fully kinetic simulation. Figure~\ref{fig:phi1A2A_2d} shows the electric potential contours in the $x=0$ m plane for the low-density (1A) and high-density (2A) cases. In these fully kinetic simulations, the sheaths are created both at the thruster exit and the downstream wall. However, because of the large local positive charge density inside the sheath, i.e., the right-hand-side of Poisson's equation (Eq.~\ref{eq:poisson}), the maximum potential in the high-density case is larger than that in the low-density case. The potential as a function of $y$-position is shown in Fig.~\ref{fig:phi1A2A_line} and indicates that the high-density case has a potential gradient (i.e., electric field, $E$) of more than twice as large as the low-density case at the edge of the beam ion plume ($2<y/R<3$).

\begin{figure}[hbt!]
    \begin{subfigure}{0.54\textwidth}
        \centering
        \includegraphics[width=\linewidth]{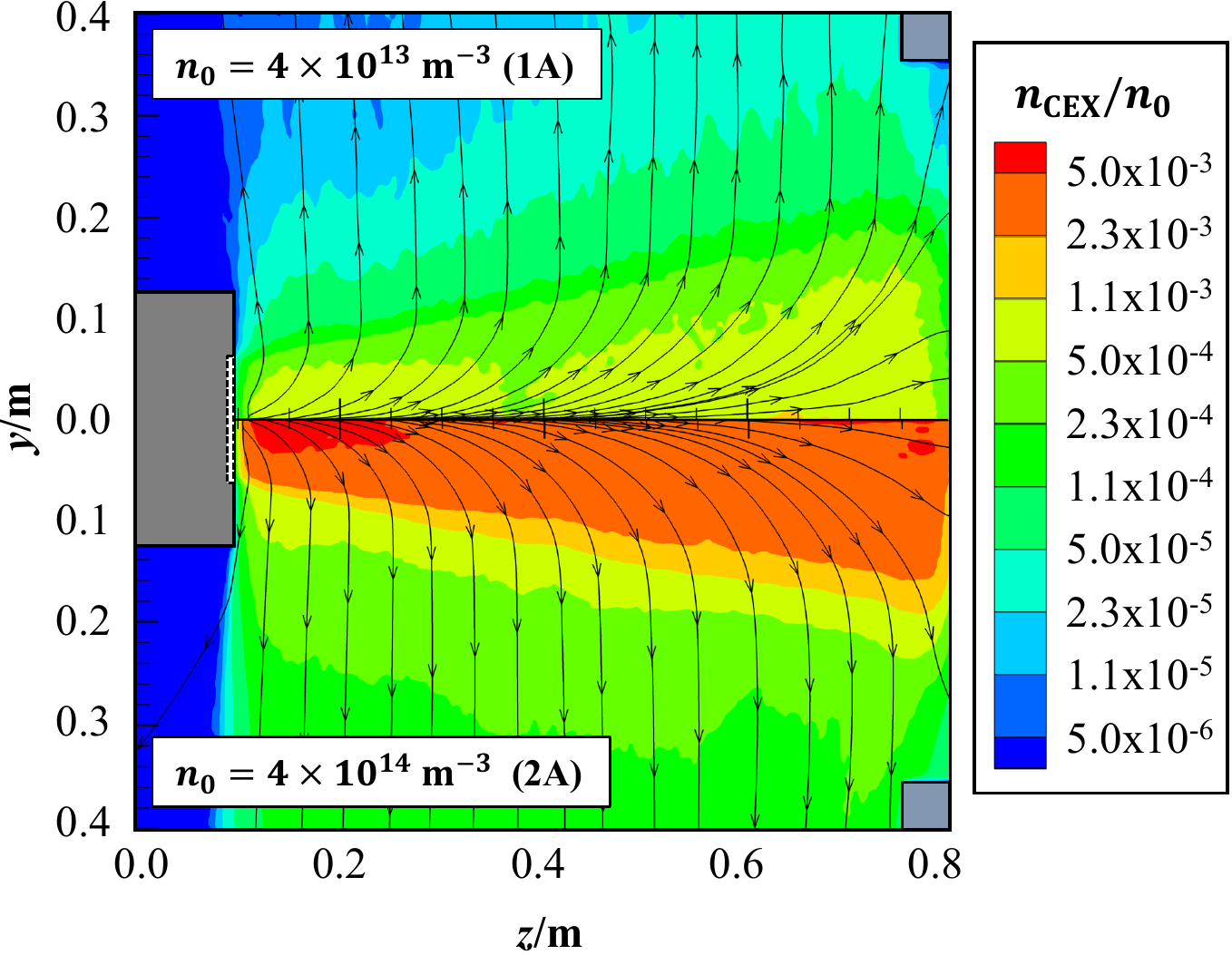}
        \caption{In the thruster center plane ($x=0$ m)} 
        \label{fig:nCEX_1A2A_2d}
    \end{subfigure}%
    \begin{subfigure}{0.46\textwidth}
        \centering
        \includegraphics[width=0.95\linewidth]{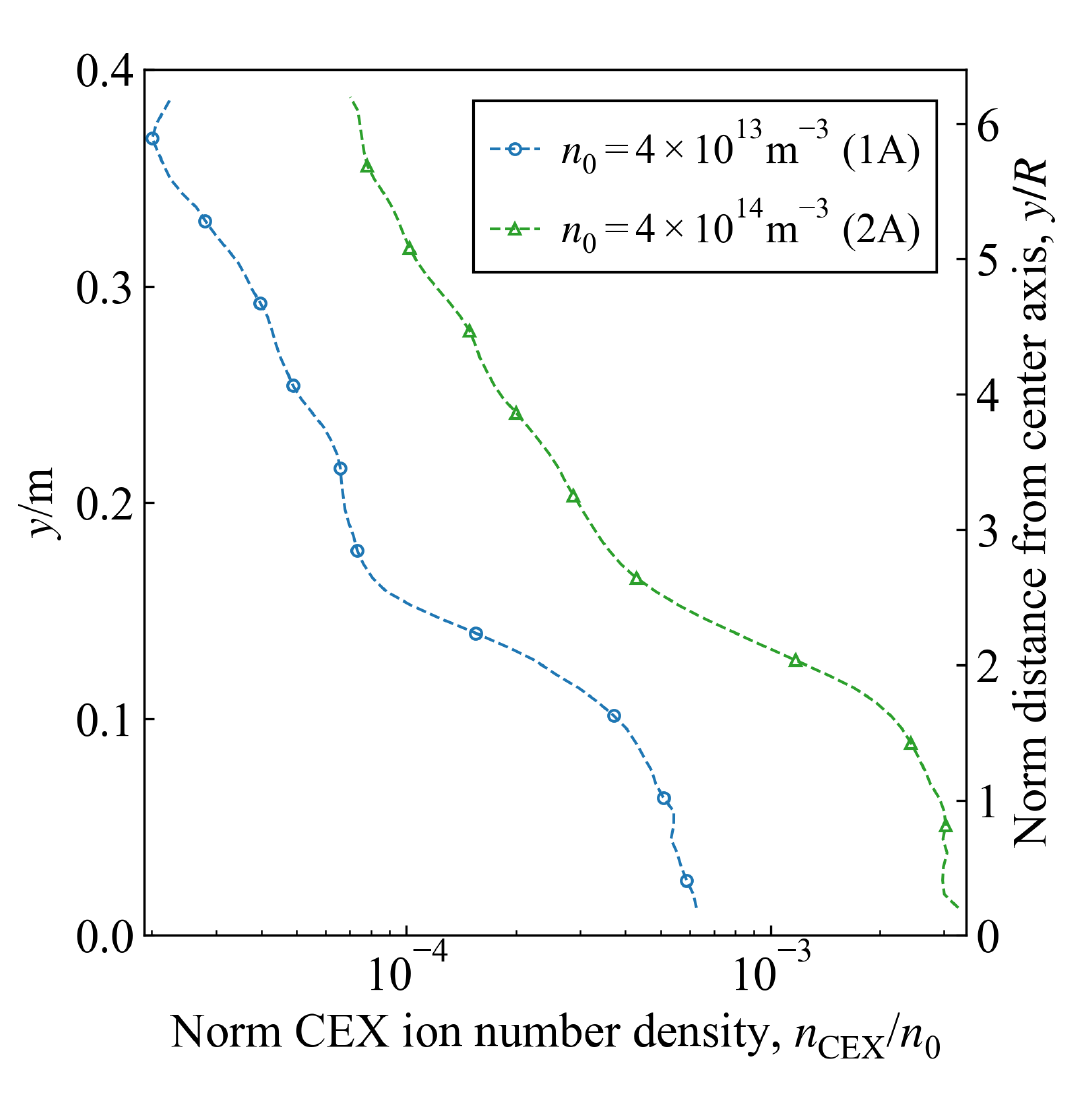}
        \caption{Along the $y$-direction ($x=0$ m, $z=0.45$ m)} 
        \label{fig:nCEX_1A2A_line}
    \end{subfigure}%
    \caption{CEX ion number density comparison between the low-density (1A) and high-density (2A) cases, where $\mathbf{\mathit{n}_0 = 4.0\times 10^{13}}$ m$^{-3}$ in 1A, $\mathbf{\mathit{n}_0 = 4.0\times 10^{14}}$ m$^{-3}$ in 2A. The streamlines shown in (a) are obtained from the CEX ion velocities. $R$ in (b) is the thruster radius of 0.0625 m.}
    \label{fig:CEXdensity_1A2A}
\end{figure}

Figure~\ref{fig:nCEX_1A2A_2d} shows the normalized number density and streamlines of CEX ions in the $x=0$ m plane, where the normalizing number densities, $\mathit{n}_0$, are different: $\mathit{n}_0 = 4.0\times 10^{13}$ m$^{-3}$ for the 1A case and $\mathit{n}_0 = 4.0\times 10^{14}$ m$^{-3}$ for the 2A case, respectively. Even after the CEX ion density is normalized by their exit ion number density, the high-density case shows an approximately 10 times larger number density. The difference can be quantitatively seen in Fig.~\ref{fig:nCEX_1A2A_line}, which shows the CEX number density in the cross-stream direction at $z=0.45$ m. The following equation expresses the production rate of CEX ions:
\begin{equation}
    \frac{dn_\mathrm{CEX}}{dt}=n_\mathrm{i}n_\mathrm{n}g\sigma_\mathrm{CEX}    
\end{equation}
where $g$ is the relative speed between an ion and a neutral particle, and $\sigma_\mathrm{CEX}$ is the CEX collision cross-section. The amount of CEX ions produced would increase with the square of the rate of increase in $n_0$ since $n_\mathrm{n} \propto n_0$ and $n_\mathrm{i} \propto n_0$ (see Table~\ref{tab:thrustercondition}). Therefore, the difference shown in Fig.~\ref{fig:nCEX_1A2A_2d} is reasonable.

\begin{figure}[hbt!]
    \centering
    \includegraphics[width=\textwidth]{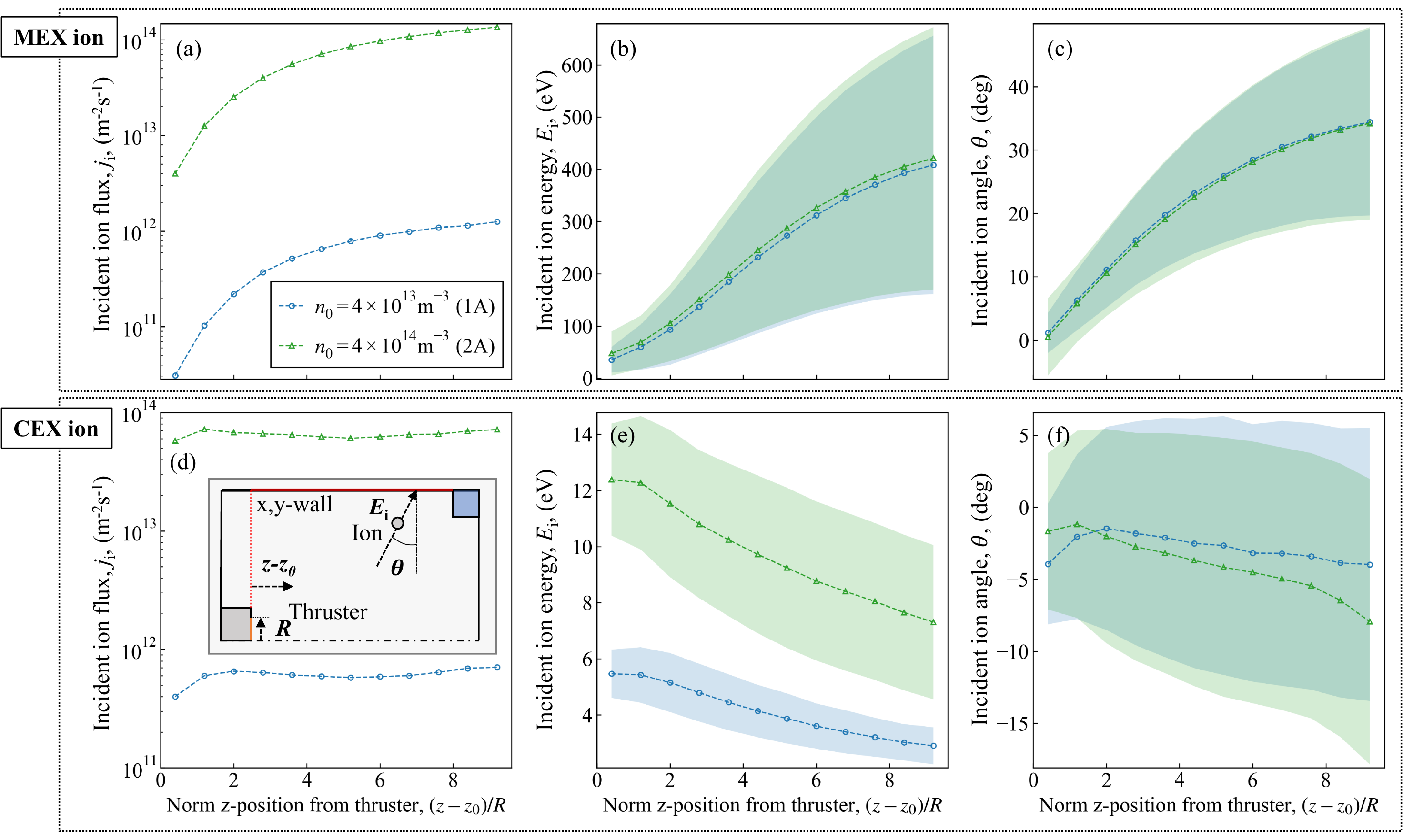}
    \caption{Particle flux, energy, and angle of the incident ions on the side walls ($x$ = 0.4 m and $y$ = 0.4 m) for the MEX ions (top) and CEX ions (bottom) in the 1A and 2A cases. On the energy and angle plots, the markers indicate the average values, and the color bands indicate the standard deviation in each bin. $z_0$ and $R$ are the thruster exit $z$-position of 0.1 m and the thruster radius of 0.0625 m, respectively.}
    \label{fig:sidewallincident_1A2A}
\end{figure}

Similar to Fig.~\ref{fig:sidewallincident_2A2B}, we sample incident ions on the side walls to compare the variation due to number density. Figure~\ref{fig:sidewallincident_1A2A} shows the particle flux, energy, and angle of incident ions on the $x$ = 0.4 m and $y$ = 0.4 m walls for MEX ions (top) and CEX ions (bottom), respectively. The wall incident flux of MEX ion shown in Fig~\ref{fig:sidewallincident_1A2A} differs by a factor of approximately 100 since the effect of density changes the collision rate between neutrals and ions, as described in the previous paragraph. In terms of the incident energy and angle of the MEX ions (Figs~\ref{fig:sidewallincident_1A2A}b, c), however, there are no differences similar to the kinetic-Boltzmann comparison (Fig.~\ref{fig:sidewallincident_2A2B}). In contrast, for the CEX ion, in addition to the difference in the flux, there is also a difference in energy. As shown in Fig.~\ref{fig:electricpotential_1A2A}, the higher the density, the greater the potential, and therefore the greater the incident energy of the CEX ions. Because of the similar potential distribution, the energy curves are also similar, and the surface incidence angles show little difference.

\subsection{Effect of Sputter Model on Carbon Backsputtering}
\label{sec:result-backsputter}

\begin{figure}[hbt!]
    \centering
    \includegraphics[width=\textwidth]{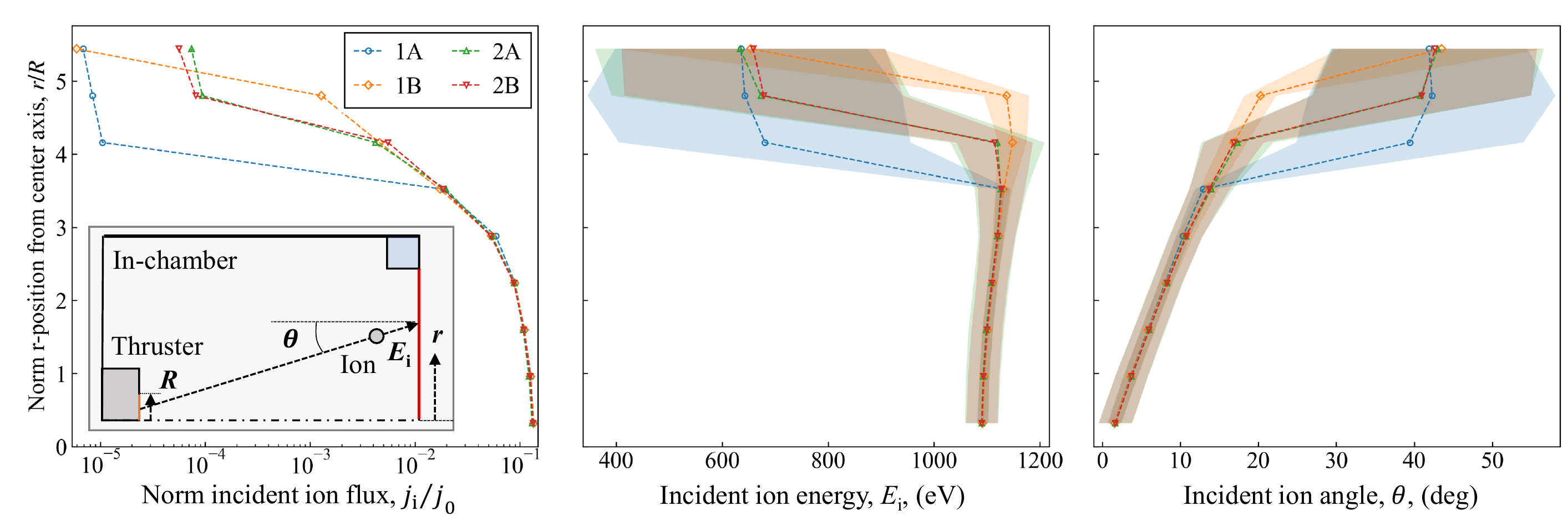}
    \caption{Particle flux, energy, and angle of the incident ions on the downstream wall ($z$ = 0.8 m), where $j_0= 1.0\times 10^{18}$ /m$^2$s in the 1A and 1B cases, $j_0= 1.0\times 10^{19}$ /m$^2$s in the 2A and 2B cases. On the energy and angle plots, the markers indicate the average values, and the color bands indicate the standard deviation in each bin. $r$ and $R$ are the distance from the thruster center axis and the thruster radius of 0.0625 m.}
    \label{fig:endwallhit}
\end{figure}

This section calculates the spatial distribution of sputtered carbon and the amount of backsputtered carbon deposited on the thruster face for the different sputtering models and ion beam currents. Sputtering is mainly caused by high-energy beam ions incident on vacuum chamber downstream walls or a beam dump. Figure~\ref{fig:endwallhit} shows the results obtained by sampling ions incident on this downstream surface. The position on the flux plot where the flux suddenly decreases is the edge of the beam ion plume ($r/R=4$), which shows the ion beam width in each case. Outside of the plume ($r/R>4$), where the flux is less than 10\% of its maximum value, the flux and energy distribution vary from case to case. Inside the beam ion plume ($r/R<4$), all cases show the same distribution. Although the CEX ions incident on the side wall depends on the electric potential model and thruster exit number density, the CEX flux and incident energy shown in Fig.~\ref{fig:sidewallincident_2A2B} is sufficiently smaller than that of beam ions shown in Fig.~\ref{fig:endwallhit}. As for MEX ions, they have high enough energy to cause sputtering, but there is no difference between the fully kinetic and Boltzmann cases, as shown in Fig.~\ref{fig:sidewallincident_2A2B}. Therefore, because the influence of electric field modeling on backsputtering is negligible, additional sputtering calculations only for the 1B and 2B cases, the Boltzmann calculations, are considered.

\begin{figure}[hbt!]
    \centering
    \includegraphics[width=0.5\linewidth]{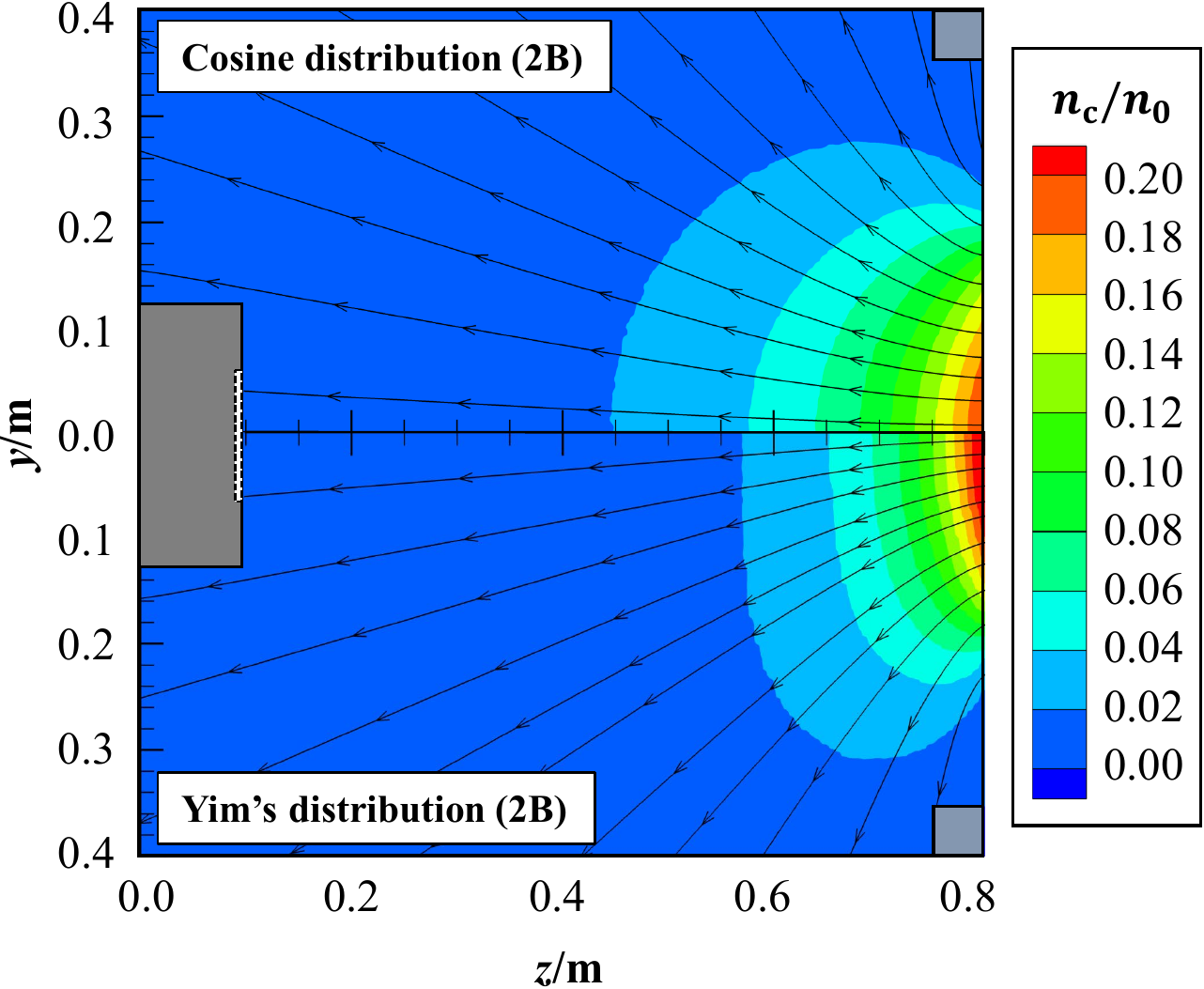}
    \caption{Number density of sputtered carbon in the thruster center plane ($x=0$ m) for the 2B case, where $n_0 = 4.0\times 10^{14}$ m$^{-3}$. Two types of sputter differential yield are tested: cosine distribution (top) and Yim's model~\cite{Yim2022-dj} (bottom). The streamlines are obtained from the carbon particle velocities.}
    \label{fig:sputterdist}
\end{figure}

\begin{figure}[hbt!]
    \begin{subfigure}{0.5\textwidth}
        \centering
        \includegraphics[width=0.9\linewidth]{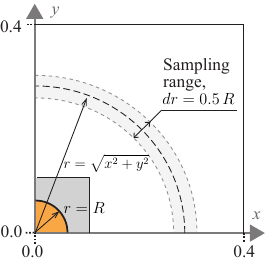}
        \caption{Schematic of sputtered carbon sampling} 
        \label{fig:sputtersample}
    \end{subfigure}%
    \begin{subfigure}{0.5\textwidth}
        \centering
        \includegraphics[width=\linewidth]{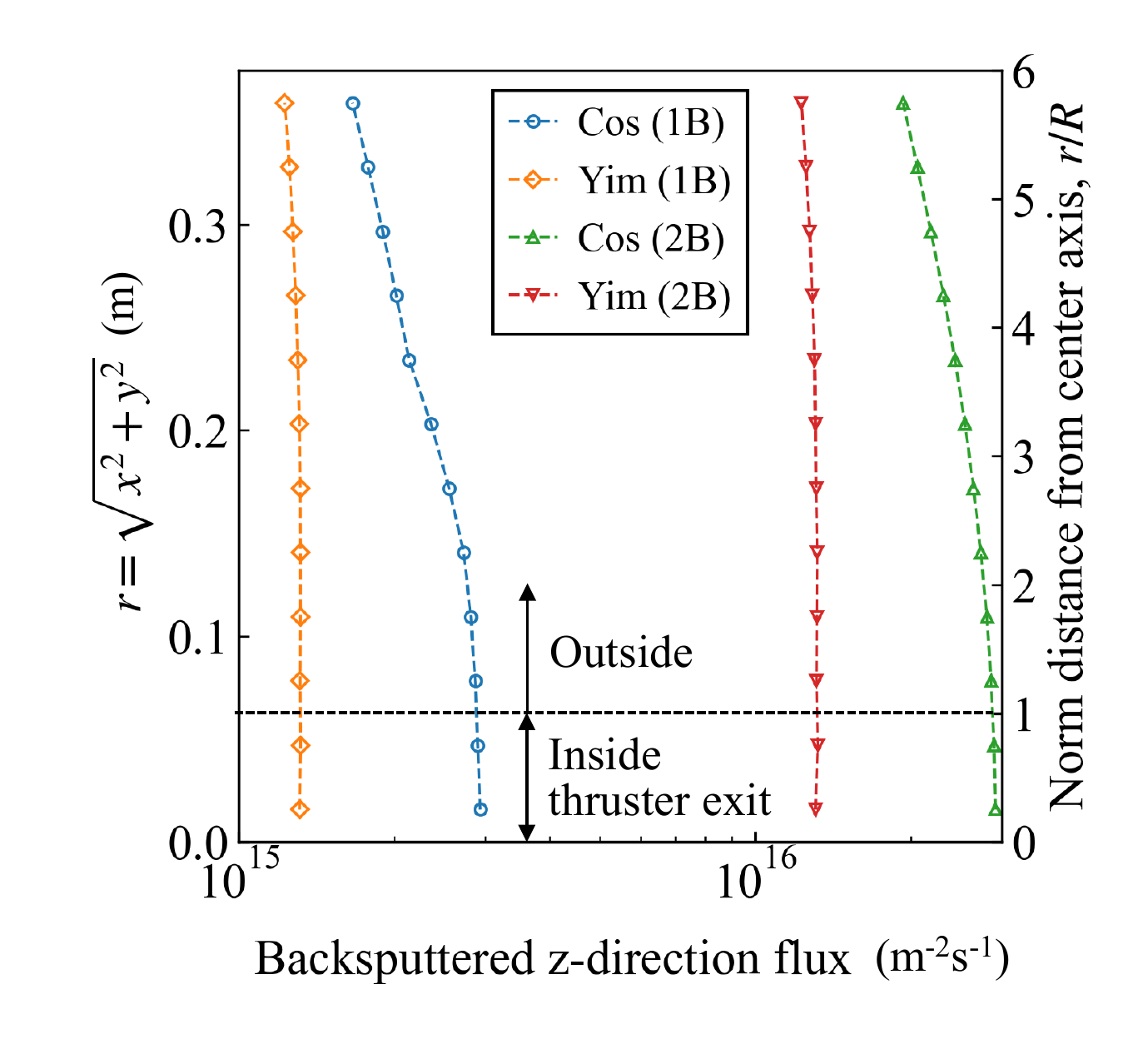}
        \caption{Particle flux as a function of radial distance} 
        \label{fig:sputterbackflux}
    \end{subfigure}%
    \caption{Sputtered carbon particle crossing the thruster face plane ($z=0.1$ m). The fluxes within $r/R<1$ are deposited on the thruster surface, where $r$ and $R$ are the distance from the thruster center axis and the thruster radius of 0.0625 m.}
    \label{fig:sputtersampling}
\end{figure}

Figure~\ref{fig:sputterdist} shows the sputtered carbon distribution on the thruster center plane ($x=0$ m) for the 2B case, where the result obtained with the cosine distribution is shown at the top, and the result by using Yim's model~\cite{Yim2022-dj} is shown at the bottom. Yim's model shows a more flattened distribution in the $y$-direction because it incorporates the experimental results that sputtered carbon has a larger polar angle than that assumed by cosine distribution, as seen in Ref~\cite{Tran2023-qk}. To obtain the backsputtered flux, we sample the sputtered carbon particles crossing the thruster face plane ($z=0.1$ m), collecting the sampled particles based on the $r$-coordinate and divided by bins with $dr=0.5R$, as shown in Fig.~\ref{fig:sputtersample}. The particles crossing within $r<R$ are deposited on the thruster surface. Figure~\ref{fig:sputterbackflux} shows the particle flux of sputtered carbon on the $z=0.1$ m plane, where it can be seen that the obtained fluxes are also flatter and smaller in Yim's model. Deposition rates of sputtered material have been mostly measured in experiments right next to the thruster in the same plane as the thruster exit~\cite{Hickman2005-ha, Gilland2016-ed}. The present simulation shows that the flux in the $r$-direction is fairly uniform in Yim's model, indicating that the deposition rate to the thruster exit can be replaced by a measurement next to the thruster.

Table~\ref{tab:backsputter} shows the carbon deposition rate on the thruster exit, which is the white area in Fig.~\ref{fig:sputterdist} and also the $r/R<1$ region in Fig.~\ref{fig:sputterbackflux}, for the low-density (1B) and high-density (2B) cases. Here, the sticking coefficient of the carbon to the thruster exit is assumed to be unity, the same as that of the chamber walls. Due to the 10 times larger beam current, the deposition particle flux is 10 times larger in the 2B case than in the 1B case. When being compared to different current densities and thruster sizes, the number flux is converted to the thickness using carbon density of 2.25 \si{g/cm^3} and then normalized by the total ion beam current~\cite{Van_Noord2005-kc}. The deposition rates per beam ampere obtained are 46 and 21 for the cosine distribution and Yim's model, respectively, for both the 1B and 2B cases, as shown in Table~\ref{tab:backsputter}, which also shows the fractions of backsputtered carbon at the side wall to the total. As mentioned in the first paragraph of this section, these fractions are very small but increase as the density increase because the number of high-energy MEX ions incident on the side walls increases as the background neutral density increases, as shown in Fig.~\ref{fig:sidewallincident_1A2A}. However, note that for even higher background pressure conditions, the effect of collisions between ions and background neutral particles on the amount of back sputtering would be larger and non-negligible.

\begin{table}[hbt!]
\caption{DSMC result of the backsputtered carbon on the thruster exit for the 1B and 2B cases.}
\centering
\begin{tabular}{cc|ccc}
\hline \hline
Case    & \bettershortstack[l]{0.5}{1.0}{Sputtered particles \\ angular distribution}
        & \bettershortstack[l]{0.5}{1.0}{Backsputtered particle \\ flux (\si{m^{-2} s^{-1}})} 
        & \bettershortstack[l]{0.5}{1.0}{Deposition rate \\/beam  amp. (\si{\micro m /kh/A})} 
        & \bettershortstack[l]{0.5}{1.0}{Particle fraction \\ from side walls, \%} \\ \hline
1B  & Cosine & $2.91\times 10^{15}$ & 46.3 & 0.011 \\
    & Yim's model  & $1.31\times 10^{15}$ & 20.9 & 0.007 \\ \hline
2B  & Cosine & $2.90\times 10^{16}$ & 46.2 & 0.103 \\
    & Yim's model  & $1.32\times 10^{16}$ & 21.0 & 0.065 \\
\hline \hline
\end{tabular}
\label{tab:backsputter}
\end{table}

Finally, we compare the deposition rates obtained in this study to that measured in experimental studies~\cite{Polk2000-dk, Hickman2005-ha, Williams2005-yn}. Because the deposition rate depends not only on the beam current but also on the chamber geometry, analytical estimation~\cite{Van_Noord2005-kc} is carried out as a function of chamber length. Therefore, we also investigate the effect of chamber length on the deposition rate by changing the $z$-direction length of the vacuum chamber between 0.8 m (original), 1.6 m (two times longer), and 3.2 m (four times longer), using the 1B case as a reference.

\begin{figure}[hbt!]
\centering
\includegraphics[width=0.5\textwidth]{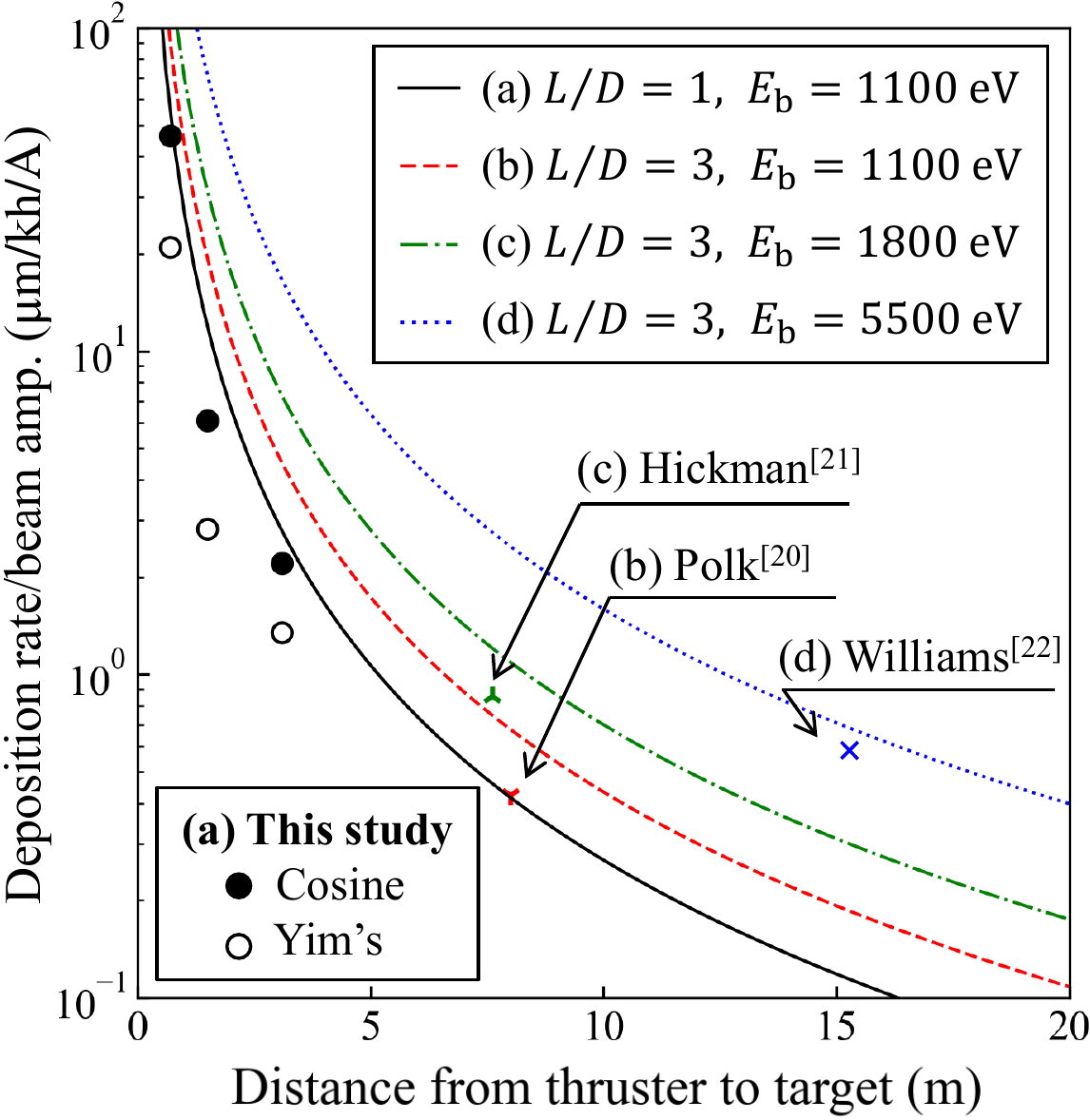}
\caption{Deposition rate per ion beam current of this study and previous experimental works~\cite{Polk2000-dk, Hickman2005-ha, Williams2005-yn}. The line plots show the analytical backsputtering rate estimation by Van Noord and Soulas~\cite{Van_Noord2005-kc}; (a) indicates this study, (b) indicates Polk~\cite{Polk2000-dk} (NSTAR), (c) indicates Hickman~\cite{Hickman2005-ha} (NEXT), and (d) indicates Williams~\cite{Williams2005-yn} (NEXIS), where $L/D$ is the aspect ratio of the vacuum chamber (length over diameter), and $E_\mathrm{b}$ is the ion beam energy.}
\label{fig:backsputcompare}
\end{figure}

Figure~\ref{fig:backsputcompare} shows the carbon deposition rate normalized by the ion beam current of this study and the experimental results as a function of the distance from the thruster exit to the downstream target or wall. Based on the formula in the previous study~\cite{Van_Noord2005-kc}, which uses the cosine distribution in sputtered angle, the deposition rate per beam current can be estimated as a function of chamber length, given the chamber aspect ratio ($L/D$) and beam ion energy ($E_\mathrm{b}$). This study uses the same coefficients of chamber geometry and the plume of the previous studies~\cite{Polk2000-dk, Hickman2005-ha, Williams2005-yn} as described in Ref.~\cite{Van_Noord2005-kc}. The letters (a)-(d) shown in Fig.~\ref{fig:backsputcompare} represent the thruster operating conditions; (a) indicates this study, (b) indicates Polk~\cite{Polk2000-dk} (NSTAR), (c) indicates Hickman~\cite{Hickman2005-ha} (NEXT), and (d) indicates Williams~\cite{Williams2005-yn} (NEXIS). Our cosine results agree with analytical line (a), which also assumes a cosine distribution. However, our deposition rates with Yim's more realistic model are lower than the estimation. In other words, the cosine distribution prediction would overestimate the carbon deposition rate. The experimental backsputtered particle distribution should be closer to the Yim distribution than the cosine distribution. The prediction lines that assume the cosine distribution exceed the experimental results, a reason that can be attributed to the difference in the angular models we have presented. 

\section{Conclusion}

In this article, we have simulated ion thruster plumes in a vacuum chamber and space configurations using the in-house multi-GPU CHAOS solver. This study applies a fully kinetic PIC-DSMC approach to three-dimensional vacuum chamber geometry without dimensional scaling for the first time. We have investigated the effect of vacuum chamber geometry, Boltzmann modeling, and plasma density on electric potential and facility plasma, which includes the ions generated by collisions between beam ions and background neutral gas. The comparison between in-space and in-chamber geometries has been carried out using fully kinetic simulations. The grounded walls on the domain boundary reduce the electric potential of the plume, and the higher background neutral number density significantly increases the CEX ion number density by 100 times, especially downstream. These differences suggest the importance of simulating ground thruster operation.

A fully kinetic approach allows the electric potential to be obtained without assuming a reference potential, which is required in the Boltzmann cases assuming quasi-neutrality and an isothermal electron temperature. Since the quasi-neutral Boltzmann simulation does not reproduce ion sheath near the thruster exit and grounded walls, a smaller electric potential is calculated than that of the fully kinetic case, resulting in non-physical CEX ion flow in the thruster off-axis region. This sheath modeling is essential, especially for simulating a ground facility surrounded by metallic walls. Further investigation on the electrical facility effect will be presented in future work. However, with respect to sputtering, this difference is insignificant because the CEX ion flux and energy are negligibly small compared to beam ions.

The higher ion number density at the thruster exit results in not only a high number density of the facility plasma but a larger incident energy of CEX ions due to the increase in electric potential inside the thruster plume. Our results predict that the maximum electric potential would increase and more facility plasma would be produced as the thruster plume density increases. The ion thrusters used in actual missions, which have about ten times larger plume density, would see even more erosion and be exposed to more contamination than our simulation results.

In addition to plasma plume modeling, we have also performed simulations of carbon backsputtered from the vacuum chamber walls. A sputtering module has been newly implemented in CHAOS, and two angular distribution models for sputtered carbon are compared. Assuming Yim's model, a semi-empirical expression based on experimental measurements, the amount of carbon backsputtered from the walls is less than half of the classical cosine distribution. Although the percentage of backsputtered carbon from the side walls is small for the geometries in this study, the fraction of Yim's model also becomes almost half of the cosine distribution. Comparisons using different chamber lengths with previous experiments and analytical models reveal that analytical models using the cosine distribution overestimate experiments, indicating the importance of an accurate angle-dependent sputtering model. Since this study demonstrates the simultaneous calculation of carbon and Xe ions, the CEX collisions between sputtered carbon and plasma will be studied in future work.

\section*{Appendix}

\subsection{Convergence Study and Parameters for Steady-state Sampling}
\label{subsec:convergence}

\begin{figure}[hbt!]
    \centering
    \includegraphics[width=\textwidth]{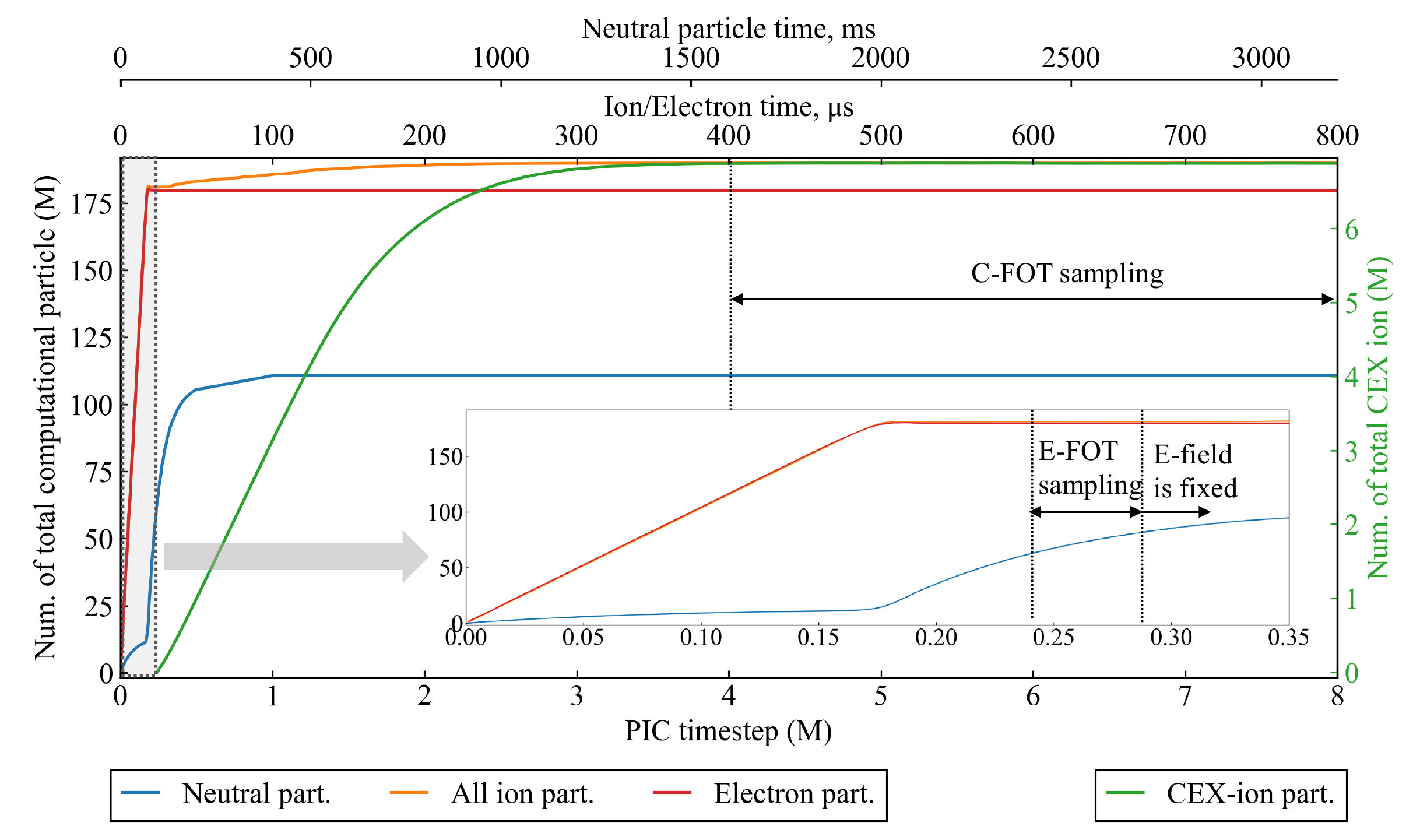}
    \caption{Total number of the computational particles during the calculation of the 2A case.}
    \label{fig:timehist}
\end{figure}

This section shows the convergence of the PIC-DSMC calculation and how we obtain a steady-state result. Figure~\ref{fig:timehist} shows the number of neutral particles, all ion, electron, and CEX ion computational particles as a function of PIC timestep. As a result of the time-slicing between the DSMC and PIC modules and species-dependent timesteps, the time scale of neutral particles and ion/electron is different, as shown at the top of the figure. We start the E-FOT sampling at 240,000 PIC timestep and fix the E-field at 280,000 PIC timestep. After the number of CEX ions reaches a steady state at four million PIC timesteps (40,000 DSMC timesteps), the C-FOT sampling, including neutral particle and ion information, starts and continues until eight million PIC timesteps (80,000 DSMC timesteps). The sampling steps for each case are shown in Table~\ref{tab:samplingsteps}.

\begin{table}[hbt!]
\caption{Number of timesteps prior to sampling and used for sampling.}
\centering
\begin{tabular}{cccccc}
\hline \hline
Simulation parameters & 0A & 1A & 2A & 1B & 2B \\ \hline
Total C-FOT steps prior to sampling & 40,000  & 30,000  & 40,000  & 100,000 & 100,000 \\
Total C-FOT steps used for sampling & 40,000  & 70,000  & 40,000  & 100,000 & 100,000 \\
Total E-FOT steps prior to sampling & 240,000 & 200,000 & 240,000 & 100,000 & 100,000 \\
Total E-FOT steps used for sampling & 40,000  & 40,000  & 40,000  & 100,000 & 100,000 \\
\hline \hline
\end{tabular}
\label{tab:samplingsteps}
\end{table}

We also show the total number of computational particles and FOTs in Table~\ref{tab:numpart}. In the high-density fully kinetic cases (0A and 2A), the number of E-FOTs and ion and computational electron particles required increases due to the small Debye length. On the other hand, since the number of neutral particles required for the calculations remains the same, the weighting factor is adjusted as shown in Table~\ref{tab:computationalparameters} to achieve a similar number of neutral particles for each calculation, thus reducing the memory cost required for these calculations.

\begin{table}[hbt!]
\caption{Steady-state computational particles and leaf node information for each case.}
\centering
\begin{tabular}{cccccc}
\hline \hline
Case ID & 0A & 1A & 2A & 1B & 2B \\ \hline
No. of comp Xe particles           & $9.9\times 10^{6}$ & $1.3\times 10^{8}$ & $1.1\times 10^{8}$ & $1.3\times 10^{8}$ & $6.5\times 10^{7}$ \\
No. of comp Xe$^+$ particles       & $1.8\times 10^{8}$ & $1.1\times 10^{7}$ & $1.9\times 10^{8}$ & $1.1\times 10^{7}$ & $5.9\times 10^{6}$ \\
No. of comp CEX Xe$^+$ particles   & $3.6\times 10^{5}$ & $8.4\times 10^{4}$ & $6.9\times 10^{6}$ & $7.5\times 10^{4}$ & $4.1\times 10^{5}$ \\
No. of comp e$^-$ particles        & $1.8\times 10^{8}$ & $1.1\times 10^{7}$ & $1.8\times 10^{8}$ & -                  & - \\
Total no. of C-FOT leaf nodes & $3.5\times 10^{5}$ & $6.8\times 10^{5}$ & $5.2\times 10^{5}$ & $5.4\times 10^{5}$ & $5.2\times 10^{5}$ \\
Total no. of E-FOT leaf nodes & $7.8\times 10^{6}$ & $5.3\times 10^{5}$ & $8.1\times 10^{6}$ & $8.9\times 10^{4}$ & $7.1\times 10^{4}$ \\
\hline \hline
\end{tabular}
\label{tab:numpart}
\end{table}

\subsection{Energy and Angle Distribution of Facility Plasma Incident on the Side Walls}
\label{subsec:sidehitEDF}

\begin{figure}[hbt!]
    \centering
    \includegraphics[width=\textwidth]{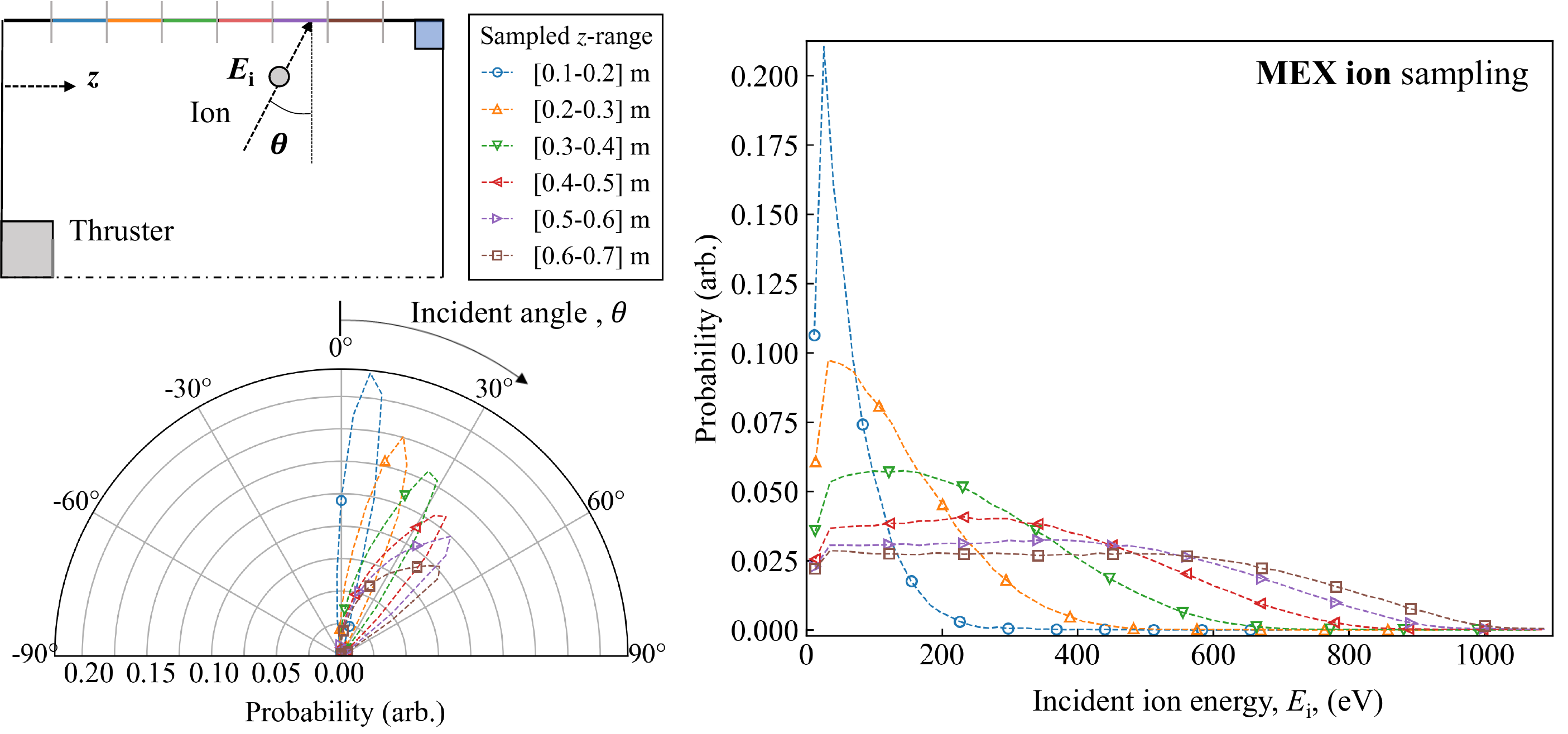}
    \caption{Angular (lower left) and energy (right) distributions of MEX ions incident on the side walls in the 2A case. The computational MEX ions are sampled at the corresponding locations shown in the upper left sketch.}
    \label{fig:sidewallEDF_MEX}
\end{figure}

\begin{figure}[hbt!]
    \centering
    \includegraphics[width=\textwidth]{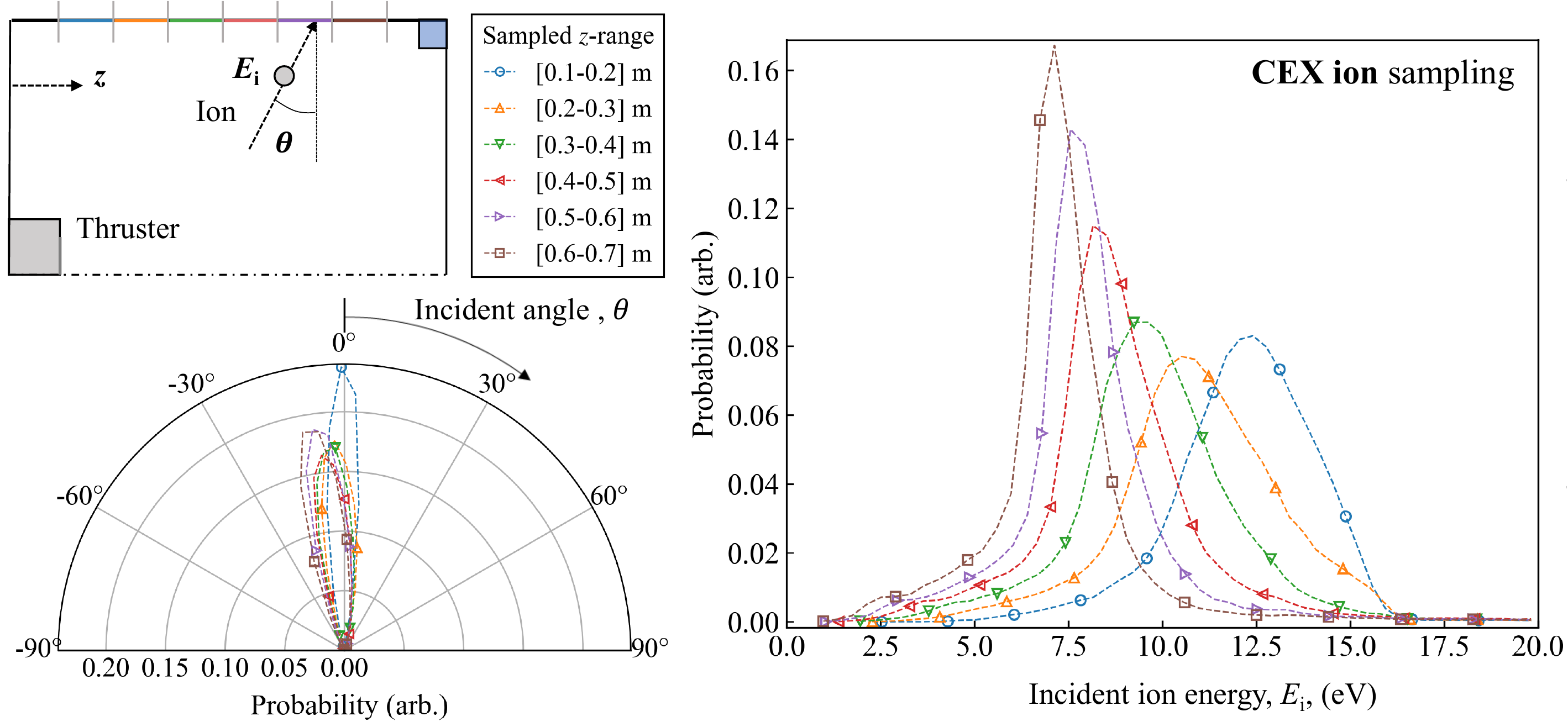}
    \caption{Angular (lower left) and energy (right) distributions of CEX ions incident on the side walls in the 2A case. The computational CEX ions are sampled at the corresponding locations shown in the upper left sketch.}
    \label{fig:sidewallEDF_CEX}
\end{figure}

This section shows the detailed energy and angle distribution for the 2A case, which is considered to be most close to an actual thruster operation condition. Figures~\ref{fig:sidewallEDF_MEX} and \ref{fig:sidewallEDF_CEX} show the energy, $E_\mathrm{i}$, and angle, $\theta$, distribution of incident ions on the side wall for MEX and CEX ions, respectively. The computational ion particles are sampled when they hit the walls, and the total number of the sampled computational particles is 2.0 million for MEX ions and 1.7 million for CEX ions. The results are divided into 6 spatial bins between $z=0.1$ m, the thruster exit position, and $z=0.7$ m, as shown in the sketch at the upper left in each figure. Looking at the MEX ions, as the incident position moves downstream, the peak energy and the spread of the distribution increase. A similar trend can be observed for the incident angles. As for CEX ions, the energy trend is opposite to the MEX ions. As the incident position moves downstream, the peak energy and the spread of the distribution decrease. There is no dependency of incident position on the incident angle on the incident position.

\section*{Acknowledgments}
This work was partially supported by NASA through the Joint Advanced Propulsion Institute, a NASA Space Technology Research Institute, grant number 80NSSC21K1118. This work used Expanse at the San Diego Supercomputer Center and Delta at the National Center for Supercomputing Applications through allocation TG-PHY220010 from the Advanced Cyberinfrastructure Coordination Ecosystem: Services \& Support (ACCESS) program, which is supported by National Science Foundation grants \#2138259, \#2138286, \#2138307, \#2137603, and \#2138296.


\bibliography{reference}

\end{document}